\documentclass[a4paper,12pt]{article}
\usepackage{epsfig}
\usepackage{color}
\usepackage[numbers,sort&compress]{natbib}

\newlength{\dinwidth}
\newlength{\dinmargin}
\begin{document}
\title{ Studying  Double Charm Decays of  $B_{u,d}$ and $B_{s}$ Mesons\\ in the MSSM with R-parity Violation }
\author{C. S. Kim$^1$\thanks{E-mail: cskim@yonsei.ac.kr},~
Ru-Min Wang$^1$\thanks{E-mail: ruminwang@cskim.yonsei.ac.kr}~ and
Ya-Dong Yang$^2$\thanks{E-mail: yangyd@iopp.ccnu.edu.cn}
\\
{\small $^1$ \it Department of Physics and IPAP, Yonsei University, Seoul
120-479,
Korea} \\
{\small $^2$ \it Institute of Particle Physics,
 Huazhong Normal University,  Wuhan 430070, P.R.China}
 }

\maketitle \vspace{0.0cm}
\begin{abstract}
\noindent Motivated by the possible large direct $CP$ asymmetry of
$\bar{B}^0_{d}\to D^{^{+}}D^{^{-}}$ decay measured by  Belle
collaboration, we investigate double charm  $B_{u,d}$ and $B_{s}$
decays in the minimal supersymmetric standard model with R-parity
violation. We derive the bounds on relevant R-parity violating
couplings from the current experimental data, which show quite
consistent measurements among relative collaborations. Using the
constrained parameter spaces, we explore R-parity violating effects
on other observables in these decays, which have not been measured
or have not been well measured yet. We find that the R-parity
violating effects on the mixing-induced $CP$ asymmetries of
$\bar{B}^0_d\to D^{^{(*)+}}D^{^{(*)-}}$ and $\bar{B}^0_s\to
D^{^{(*)+}}_sD^{^{(*)-}}_s$ decays could be very large, nevertheless
the R-parity violating  effects on the direct $CP$ asymmetries could
not be large enough to explain the large direct $CP$ violation of
$\bar{B}^0_{d}\to D^{^{+}}D^{^{-}}$ from Belle. Our results could be
used to probe  R-parity violating effects and will correlate with
searches for  direct R-parity violating signals in future
experiments.

\vspace{0.5cm} \noindent {PACS Numbers:  12.60.Jv,
  12.15.Ji,  13.25.Hw, 14.40.Lb }

\end{abstract}

\newpage
\section{Introduction}

Double charm decays of $B_{u,d}$ and $B_{s}$  provide us with a rich
field to study $CP$ violation and final-state interactions as well
as to extract information of Cabibbo-Kobayashi-Maskawa (CKM)
elements. $CP$ asymmetries (CPAs) in these decays play important
roles in testing the Standard Model (SM) as well as exploring new
physics (NP) \cite{Sanda:1996pm,Xing:1999yx}.

Double charm decays, $\bar{B}^0_{d}\to D^{^{(*)+}}D^{^{(*)-}}$,
$B^-_{u}\to D^{^{(*)0}}D^{^{(*)-}}$ and $\bar{B}^0_{s}\to
D^{^{(*)+}}_sD^{^{(*)-}}$,  are dominated by color-allowed tree
$b\to c\bar{c}d$ transition, but involve small penguin pollution
from $b\to u\bar{u}d$ transition carrying a different weak phase.
The latter contributions lead to  direct CPAs, which are
very small (about the order of $10^{-2}$) in the SM.
If penguin corrections are neglected, the SM predictions for the
direct CPAs  would be zero.
 It is interesting to note that
both B{\footnotesize A}B{\footnotesize AR} and Belle have measured
the direct CPA in $B^{0}_{d}\to D^{+}D^{-}$ decay
\begin{eqnarray}
\mathcal{C}(B^0_d,\bar{B}^0_d\to D^+D^-)&=&\left\{\begin{array}{ll}
 -0.91\pm0.23\pm0.06 &\mbox{(Belle \cite{Fratina:2007zk})},\\
-0.07\pm0.23\pm0.03 &\mbox{(B{\footnotesize A}B{\footnotesize AR}
\cite{:2008aw})},
\end{array}\right.
\label{eq:DDdircp}
\end{eqnarray}
respectively. One would find the difference between the two measurements is
\begin{equation}
\Delta \mathcal{C}=0.84\pm0.32,
\end{equation}
{\it i.e}, the difference is as large as $2.7\sigma$.  So far, such
a large direct CPA has not been observed  in
the other measurements of
 $\bar{B}^0_{d}\to D^{^{(*)+}}D^{^{(*)-}}$,
$B^-_{u}\to D^{^{(*)0}}D^{^{(*)-}}$ and $\bar{B}^0_{s}\to
D^{^{(*)+}}_sD^{^{(*)-}}$ decays \cite{:2008aw,Miyake:2005qb,Aushev:2004uc,Aubert:2007rr,Vervink:2008dv,Aubert:2006ia,Abe:2007sk,Majumder:2005gy},
which involve the same quark level weak decays. If the large CP
violation in $\bar{B}^0_d\to D^+D^-$ from Belle is true, it would
establish the presence of NP.
At present one cannot conclude the presence of NP in those decays. Equivalently one also cannot take
the CPAs are in agreement with the SM expectations.
Recently the large direct CPA in $\bar{B}^0_d\to D^+D^-$ has been
investigated with possible NP scenarios, such as
unparticle interaction \cite{Zwicky:2007vv} and the NP  effects in
electroweak penguin sector \cite{Fleischer:2007zn,Gronau:2008ed}, and so on.

In this paper, we would like to investigate  $B_{u,d}$ and $B_{s}$ double charm decays systematically in
 the minimal supersymmetric standard model (MSSM) \cite{MSSM1,MSSM2} with R-parity violation \cite{Weinberg:1981wj,SUSY}.
In the literature,
the possible appearance of the R-parity violating
(RPV) couplings \cite{Weinberg:1981wj,SUSY}, which violate the
lepton and/or baryon number conservations, has gained full attention
in searching for supersymmetry \cite{report,allanach}.
The effects of supersymmetry with R-parity violation in $B$ meson
decays have been extensively investigated, for instance in Refs.
\cite{RPVstudy1,RPVstudy2,RPVstudy3,RPVstudy4}. In our work,
 twenty-four double charm decays
$\bar{B}^0_{d}\to D^{^{(*)+}}D^{^{(*)-}}_{_{(s)}}$,
$\bar{B}^-_{u}\to D^{^{(*)0}}D^{^{(*)-}}_{_{(s)}}$ and
$\bar{B}^0_{s}\to D^{^{(*)+}}_{_{s}}D^{^{(*)-}}_{_{(s)}}$
are studied in the RPV MSSM.  For simplicity we employ  naive factorization \cite{NF}  for the hadronic dynamics, which is
expected to be reliable for the color-allowed amplitudes, which are dominant contributions in those double charm decays.

The color-allowed tree level dominated decays of $b\to c\bar{c}d$, $i.e.$
$\bar{B}^0_{d}\to D^{^{(*)+}}D^{^{(*)-}}$,  $B^-_{u}\to
D^{^{(*)0}}D^{^{(*)-}}$ and $\bar{B}^0_{s}\to
D^{^{(*)+}}_sD^{^{(*)-}}$,  involve the same set of  RPV coupling
constants. For these processes, besides the CPA in $\bar{B}^{0}_d\to
D^{+}D^{-}$,  a few other observables in
 $B_{u,d}\to D^{(*)}D^{(*)}$
 have  been already measured by B{\footnotesize A}B{\footnotesize AR} and Belle
collaborations
\cite{Fratina:2007zk,:2008aw,Aubert:2006ia,Abe:2007sk,Majumder:2005gy,Abe:2002zy,
Miyake:2005qb,Aushev:2004uc,Aubert:2007rr,Vervink:2008dv}.
To derive constraints on the relevant RPV couplings, we will choose a set of data
from the aforementioned measurements which have quite high consistency
between the measurements of B{\footnotesize A}B{\footnotesize AR}
and Belle. Then,  using the constrained RPV coupling parameter
spaces, we predict the RPV effects on the other observables in
$\bar{B}^0_{d}\to D^{^{(*)+}}D^{^{(*)-}}$, $B^-_{u}\to
D^{^{(*)0}}D^{^{(*)-}}$ and $\bar{B}^0_{s}\to
D^{^{(*)+}}_{s}D^{^{(*)-}}$ decays, whose
measurements from B{\footnotesize A}B{\footnotesize AR} and Belle
are not compatible within $2\sigma$ error range, and/or which have
not been measured yet.
One of our goals is to see how large the direct CPA of
$\bar{B}^0_d\to D^+D^-$ can be within the constrained parameter spaces. We find that the lower limit
of $\mathcal{C}(B^0_d,\bar{B}^0_d\to D^{+}D^{-})$ could
be just slightly
decreased  by the RPV couplings, and the RPV effects on this quantity are not large enough to
 explain the large direct $CP$ violation from Belle,  although the mixing-induced CPAs of
$\bar{B}^0_{d}\to D^{^{(*)+}}D^{^{(*)-}}$ are very sensitive to the RPV couplings.

Decays $\bar{B}^0_{d}\to D^{^{(*)+}}D^{^{(*)-}}_{s}$, $B^-_{u}\to
D^{^{(*)0}}D^{^{(*)-}}_{s}$ and $\bar{B}^0_{s}\to
D^{^{(*)+}}_{s}D^{^{(*)-}}_{s}$  are governed by the $b\to
c\bar{c}s$ transition at the quark level, and also involve the same
set of RPV coupling constants. They have similar properties to
$\bar{B}^0_{d}\to D^{^{(*)+}}D^{^{(*)-}}$, $B^-_{u}\to
D^{^{(*)0}}D^{^{(*)-}}$ and $\bar{B}^0_{s}\to
D^{^{(*)+}}_{s}D^{^{(*)-}}$ decays, nevertheless the penguin effects
are less Cabibbo-suppressed. For these decays, most branching ratios and
one  longitudinal polarization   have been measured
\cite{Aubert:2006nm,Aubert:2003jj,Aubert:2005xu,Gibaut:1995tu,
Bortoletto:1991kz,Ahmed:2000ad,Bortoletto:1990fx,Albrecht:1991pa}.
We will take the same strategy as the one for $b\to c\bar{c}d$ decays to constrain
relevant RPV couplings and estimate RPV effects in these decays. We
find that RPV couplings could significantly affect the CPAs of these
decays, and could flip their signs.

Our paper is organized as follows: In Sec. 2,  we briefly introduce
the theoretical framework for the double charm $B_{u,d}$ and $B_{s}$ decays
in the RPV MSSM, and we tabulate all the theoretical input
parameters. In Sec. 3, we deal with the numerical results and our discussions.
At first, we give the SM predictions with full uncertainties of the input
parameters. Then, we derive
the constrained parameter spaces
which satisfy all the experimental data with  high consistency between
different collaborations.  Finally, we predict the RPV effects on other
quantities, which have not been measured or have not been well
measured yet. Section 4 contains our summary.

\section{Theoretical Framework}

\subsection{Decay amplitudes in the SM}

 In the SM, the low energy effective Hamiltonian for
$\Delta B=1$ transition at a scale $\mu$  is given by
\cite{Buchalla:1995vs}
 \begin{eqnarray}
 \mathcal{H}^{\rm SM}_{\rm eff}&=&\frac{G_F}{\sqrt{2}}\sum_{p=u, c}
 \lambda_p \Biggl\{C_1Q_1^p+C_2Q_2^p
 +\sum_{i=3}^{10}C_iQ_i+C_{7\gamma}Q_{7\gamma}
 +C_{8g}Q_{8g} \Biggl\}+ h.c.,
 \label{HeffSM}
 \end{eqnarray}
here  $\lambda_p=V_{pb}V_{pq}^* $ for $b \to q$ transition $(p\in
\{u,c\},q\in \{d,s\})$. The detailed definition of the effective
Hamiltonian can be found in \cite{Buchalla:1995vs}.

It is empirically observed that naive factorization \cite{NF}
 still works reasonably well in the color-allowed double charm $B_{u,d}$ and $B_{s}$ decay
processes. We will describe the $B\to D^{^{(*)}}D^{^{(*)}}_{_{q}}$
decay amplitudes within the naive factorization approximation in
this paper. Under the naive factorization approximation, the
factorized matrix elements are given by
\begin{eqnarray}
A_{[BD^{^{(*)}},D^{^{(*)}}_{_{q}}]}\equiv\left<D^{^{(*)}}_{_{q}}
|\bar{q}\gamma^\mu(1-\gamma_5)c|0\right>\left<D^{^{(*)}}|\bar{c}\gamma_\mu(1-\gamma_5)b|B\right>.
\end{eqnarray}
Decay constants and form factors \cite{Wirbel:1985ji,Neubert:1991xw}
 are usually defined as
\begin{eqnarray}
\langle D_q(p_{_{D_q}})|\bar{q}\gamma^\mu\gamma_5c|0\rangle &=& -if_{_{D_q}}p^\mu_{_{D_q}},\\
\langle D^*_q(p_{_{D^*_q}})|\bar{q}\gamma^\mu c|0\rangle &=& f_{_{D^*_q}}p^\mu_{_{D^*_q}},\\
\langle D(p_{_{D}})|\bar{c}\gamma_{\mu}b|B (p_{_{B}})\rangle
&=&\left[(p_{_{B}}+p_{_{D}})_\mu-\frac{m^2_B-m^2_{_{D}}}{q^2}q_\mu\right]
F_1(q^2)+\frac{m^2_B-m^2_{_{D}}}{q^2}q_\mu F_0(q^2),\\
\langle D^*(p_{_{D^*}},\varepsilon^{\ast})|\bar{c}\gamma_{\mu}b|B
(p_{_{B}})\rangle &=& \frac{2V(q^2)}{m_B+m_{_{D^*}}}
\epsilon_{\mu\nu\alpha\beta}\varepsilon^{\ast\nu}p_{_B}^{\alpha}p_{_{D^*}}^{\beta},\\
\langle
D^*(p_{_{D^*}},\varepsilon^{\ast})|\bar{c}\gamma_{\mu}\gamma_5b|B
(p_{_{B}})\rangle
&=&i\left[\varepsilon_{\mu}^\ast(m_B+m_{_{D^*}})A_1(q^2)
-(p_{_B}+p_{_{D^*}})_{\mu}({\varepsilon^\ast}\cdot{p_{_B}})\frac{A_2(q^2)}
{m_B+m_{_{D^*}}}\right]\nonumber \\
&&-iq_{\mu}({\varepsilon^\ast}\cdot{p_{_B}})\frac{2m_{_{D^*}}}{q^2}
[A_3(q^2)-A_0(q^2)],
\end{eqnarray}
with $q=p_B-p_{D^{^{(*)}}}$. Then we can express
$A_{[BD^{^{(*)}},D^{^{(*)}}_{_{q}}]}$ in terms of decay constants
and form factors as follows
\begin{eqnarray}
 A_{[BD^{^{(*)}},D^{^{(*)}}_{_{q}}]}=\left
 \{\begin{array}{ll}if_{D_q}(m_B^2-m^2_{D})F_0(m^2_{D_q}) &
(DD_{q}),
\\2f_{D^{^{*}}_{_{q}}}m_B|p_c|F_1(m^2_{D^{^{*}}_{_{q}}})
& (DD^{*}_{q}),
 \\-2f_{D_{_{q}}}m_B|p_c|A_0(m^2_{D_{_{q}}})
&(D^{*}D_{q}),
 \\-i f_{D^{^{*}}_{_{q}}}m_{D^{^{*}}_{_{q}}}\biggl[
(\varepsilon_{D^{^{*}}}^{\ast}\cdot\varepsilon_{D^{^{*}}_{_{q}}}^{\ast})
(m_{B}+m_{D^{^{*}}})A_1(m_{D^{^{*}}_{_{q}}}^2)\\
\hspace{2cm}-(\varepsilon_{D^{^{*}}}^{\ast}\cdot
p_{D^{^{*}}_{_{q}}})(\varepsilon_{D^{^{*}}_{_{q}}}^{\ast}\cdot
p_{D^{^{*}}})\frac{2A_2(m^2_{D^{^{*}}_{q}})}{m_{B}+m_{D^{^{*}}}}\biggr.\\\hspace{2cm}
\left.+i\epsilon_{\mu\nu\alpha\beta}\varepsilon_{D^{^{*}}_{_{q}}}^{\ast\mu}
\varepsilon_{D^{^{*}}}^{\ast\nu}p_{D^{^{*}}_{_{q}}}^{\alpha}p_{D^{^{*}}}^{\beta}
\frac{2V(m^2_{D^{^{*}}_{q}})}{m_{B}+m_{D^{^{*}}}}\right]
&(D^*D^*_{q}).
 \end{array}
 \right.
 \end{eqnarray}

Decays $B\to D^{^{(*)}}D^{^{(*)}}_{_{q}}$  may occur through both
tree level and loop induced (penguin) quark diagrams, and the SM decay
amplitudes within the naive factorization are given as
\begin{eqnarray}
\mathcal{M}^{\rm SM}(B\to
D^{^{(*)}}D^{^{(*)}}_{_{q}})=\frac{G_F}{\sqrt{2}}\left(\lambda_c
a_1^c +\sum_{p=u,c}\lambda_p\left[a_4^p+a_{10}^p+\xi(a_6^p+a_8^p)\right]
\right)
 A_{[BD^{^{(*)}},D^{^{(*)}}_{_{q}}]},\label{AM}
\end{eqnarray}
where the coefficients
$a^{p}_i=\left(C_i+\frac{C_{i\pm1}}{N_c}\right)+P^{p}_i$
with the upper (lower) sign applied when $i$ is odd (even), and
$P^p_i$ account for penguin contractions.  The factorization
parameter $\xi$ in Eq. (\ref{AM}) arises from the transformation of
$(V-A)(V+A)$ currents into $(V-A)(V-A)$ ones for the penguin
operators $Q_5,\cdots,Q_8$, and it depends on properties of the
final-state mesons
\begin{eqnarray}
\xi&=&\left\{\begin{array}{cl}
+\frac{2m^2_{D_q}}{(\bar{m}_c+\bar{m}_q)(\bar{m}_b-\bar{m}_c)} &~~\mbox{($DD_q$)},\\
0 &~~\mbox{($DD^*_q$)},\\
-\frac{2m^2_{D_q}}{(\bar{m}_c+\bar{m}_q)(\bar{m}_b+\bar{m}_c)} &~~\mbox{($D^*D_q$)},\\
0 &~~\mbox{($D^*D^*_q$)}.\\
\end{array}\right.
\end{eqnarray}
For the penguin contractions, we will consider not only QCD and
electroweak penguin operator contributions but also
contributions from the electromagnetic and chromomagnetic dipole
operators. $P^p_i$  are given as follows
\begin{eqnarray}
P_1^c &=&0, \nonumber\\
P_4^p&=&\frac{\alpha_s}{9\pi}\left\{C_1\left[
\frac{10}{9}-G_{D^{(*)}_q}(m_p)\right]-2F_1C^{eff}_{8g}\right\},\nonumber\\
P_6^p&=&\frac{\alpha_s}{9\pi}\left\{C_1\left[
\frac{10}{9}-G_{D^{(*)}_q}(m_p)\right]-2F_2C^{eff}_{8g}\right\},\nonumber\\
P_8^p&=&\frac{\alpha_e}{9\pi}\frac{1}{N_c}\left\{(C_1+N_cC_2)\left[
\frac{10}{9}-G_{D^{(*)}_q}(m_p)\right]-3F_2C^{eff}_{7\gamma}\right\},\nonumber\\
P_{10}^p&=&\frac{\alpha_e}{9\pi}\frac{1}{N_c}\left\{(C_1+N_cC_2)\left[
\frac{10}{9}-G_{D^{(*)}_q}(m_p)\right]-3F_1C^{eff}_{7\gamma}\right\},\label{PiF}
\end{eqnarray}
where the penguin loop-integral function $G_{D^{(*)}_q}(m_p)$ is given by
\begin{eqnarray}
G_{D^{(*)}_q}(m_p)&=&\int^1_0 du  G(m_p,k) \Phi_{D^{(*)}_q}(u),\\
 G(m_p,k)&=&-4\int^1_0 dx
x(1-x)\mbox{ln}\left[\frac{m_p^2-k^2x(1-x)}{m^2_b} -i \epsilon\right],
\end{eqnarray}
with the penguin momentum transfer
$k^2=m^2_c+\bar{u}(m^2_b-m^2_c-m^2_{M_2})+\bar{u}^2m^2_{M_2}$, where $\bar u \equiv 1-u$. In
the function $G_{D^{(*)}_q}(m_p)$, we have used a $D^{(*)}_q$
meson-emitting distribution amplitude
$\Phi_{D^{(*)}_q}(u)=6u(1-u)[1+a_{D^{(*)}_q}(1-2u)]$,
in stead of keeping $k^2$ as a free parameter as usual.
The constants $F_1$ and $F_2$ in Eq. (\ref{PiF}) are defined by
\begin{eqnarray}
F_1&=&\left\{\begin{array}{ll}
\int^1_0 du  \Phi_{D_q}(u)\frac{m_b}{m_b-m_c}\frac{m^2_b-um^2_{D_q}-2m^2_c+m_bm_c}{k^2} &~~\mbox{($DD_q$)},\\
\int^1_0 du  \Phi_{D^*_q}(u)\frac{m_b}{k^2}\left(\bar{u}m_b+\frac{2um_{D^*_q}}{m_b-m_c}\epsilon_2^*\cdot p_{1} -um_c\right)
 &~~\mbox{($DD^*_q$)},\\
\int^1_0 du  \Phi_{D_q}(u)\frac{m_b}{m_b+m_c}\frac{m^2_b-um^2_{D_q}-2m^2_c-m_bm_c}{k^2} &~~\mbox{($D^*D_q$)},\\
\int^1_0 du  \Phi_{D^*_q}(u)\frac{m_b}{k^2}\left(\bar{u}m_b+\frac{2um_{D^*_q}}{m_b+m_c}\epsilon_2^*\cdot p_{1} +um_c\right)
  &~~\mbox{($D^*D^*_q$)},\\
\end{array}\right.\\
F_2&=&\left\{\begin{array}{ll}
\int^1_0 du  \Phi_{D_q}(u)\frac{m_b}{k^2}[\bar{u}(m_b-m_c)+m_c] &~~\mbox{($DD_q$)},\\
0 &~~\mbox{($DD^*_q$)},\\
\int^1_0 du  \Phi_{D_q}(u)\frac{m_b}{k^2}[\bar{u}(m_b+m_c)-m_c] &~~\mbox{($D^*D_q$)},\\
0  &~~\mbox{($D^*D^*_q$)},\\
\end{array}\right.
\end{eqnarray}
where $\epsilon_{2L}^*\cdot p_{1}\approx
(m_b^2-m^2_{M^*_q}-m_c^2)/(2m_{M^*_q})$ and $\epsilon_{2T}^*\cdot
p_{1}=0$ for $B\to D^*D^*_q$ decays.

\subsection{Decay amplitudes of  the RPV contributions}

In the RPV MSSM, in terms of the RPV superpotential
\cite{Weinberg:1981wj}, we can obtain the relative RPV effective
Hamiltonian for $B\to D^{^{(*)}}D^{^{(*)}}_{_{q}}$ decays as
following
\begin{eqnarray}
\mathcal{H}^{\rm RPV}_{\rm eff}&=&\sum_n\frac{\lambda''_{ikn}
\lambda''^*_{jln}}{2m^2_{\tilde{d}_{n}}}\eta^{-4/\beta_0}
\left[-(\bar{d}_k\gamma^\mu P_Ru_j)_1 (\bar{u}_i\gamma_\mu
P_Rd_l)_1+(\bar{d}_k\gamma^\mu P_Ru_j)_8 (\bar{u}_i\gamma_\mu
P_Rd_l)_8\right]\nonumber\\
&+&\sum_i\frac{\lambda'_{ijk}
\lambda'^*_{inl}}{m^2_{\tilde{e}_{iL}}}
\eta^{-8/\beta_0}(\bar{d}_kP_Lu_j)_1(\bar{u}_n P_Rd_l)_1+h.c.,
\end{eqnarray}
where $P_L=\frac{1-\gamma_5}{2},P_R=\frac{1+\gamma_5}{2},
\eta=\frac{\alpha_s(m_{\tilde{f}})}{\alpha_s(m_b)}$ and
$\beta_0=11-\frac{2}{3}n_f$. The subscripts 1 and 8 of the currents
represent the currents in the color singlet and octet, respectively.
The coefficients $\eta^{-4/\beta_0}$ and $\eta^{-8/\beta_0}$ are due
to the running from  sfermion mass scale $m_{\tilde{f}}$ (assumed as
100 GeV) down to  $m_b$ scale. Since it is usually assumed in
phenomenology for numerical display that only one sfermion
contributes at one time, we neglect the mixing between the operators
when we use the renormalization group equation (RGE) to run
$\mathcal{H}^{\rm RPV}_{\rm eff}$ down to the low scale.

The decay amplitudes of RPV contributions to $B\to
D^{^{(*)}}D^{^{(*)}}_{_{q}}$ are given by
\begin{eqnarray}
\mathcal{M}^{\rm RPV}(B\to D^{^{(*)}}D^{^{(*)}}_{_{q}})&=&
\Lambda''\left(-1+\frac{1}{N_C}\right)
\left<D^{^{(*)}}_{_{q}}|\bar{q}\gamma_\mu(1+\gamma_5)c|0\right>
\left<D^{^{(*)}}|\bar{c}\gamma_\mu(1+\gamma_5)b|B\right>\nonumber\\
&&+ 2\Lambda'\left<D^{^{(*)}}_{_{q}}|\bar{q}(1-\gamma_5)c|0\right>
\left<D^{^{(*)}}|\bar{c}(1+\gamma_5)b|B\right>,\\
&=&\left \{\begin{array}{ll}
 \left[-\Lambda''\left(-1+\frac{1}{N_C}\right)
 +\xi\Lambda'\right] A_{[BD,D_{_{q}}]} \hspace{0.5cm}&
(DD_{q}),
\\ \left[\Lambda''\left(-1+\frac{1}{N_C}\right)\right] A_{[BD,D^*_{_{q}}]}
 &(DD^{*}_{q}),
 \\\left[\Lambda''\left(-1+\frac{1}{N_C}\right)
 -\xi\Lambda'\right] A_{[BD^*,D_{_{q}}]}
&(D^{*}D_{q}),
 \\\left[\Lambda''\left(-1+\frac{1}{N_C}\right)\right] A'_{[BD^*,D^*_{_{q}}]}
 &(D^*D^{*}_{q}),
 \end{array}
 \right.
\end{eqnarray}
where
$\Lambda''\equiv\eta^{-4/\beta_0}\frac{\lambda''^*_{232}\lambda''_{212}}{8m^2_{\tilde{s}}}
\left(\frac{\lambda''^*_{231}\lambda''_{221}}{8m^2_{\tilde{d}}}\right)$
and $\Lambda'\equiv
\eta^{-8/\beta_0}\sum_i\frac{\lambda'^*_{i23}\lambda'_{i21}}{8m^2_{\tilde{e}_{iL}}}
$
$\left(\frac{\lambda'^*_{i23}\lambda'_{i22}}{8m^2_{\tilde{e}_{iL}}}\right)$
for $q=d~(q=s)$. $A'_{[BD^*,D^*_{_{q}}]}$ is defined by
\begin{eqnarray}
A'_{[BD^*,D^*_{_{q}}]}&\equiv&i
f_{D^{^{*}}_{_{q}}}m_{D^{^{*}}_{_{q}}}\biggl[
(\varepsilon_{D^{^{*}}}^{\ast}\cdot\varepsilon_{D^{^{*}}_{_{q}}}^{\ast})
(m_{B}+m_{D^{^{*}}})A_1(m_{D^{^{*}}_{_{q}}}^2)\nonumber\\
&&-(\varepsilon_{D^{^{*}}}^{\ast}\cdot
p_{D^{^{*}}_{_{q}}})(\varepsilon_{D^{^{*}}_{_{q}}}^{\ast}\cdot
p_{D^{^{*}}})\frac{2A_2(m^2_{D^{^{*}}_{q}})}{m_{B}+m_{D^{^{*}}}}
\biggr.\left.-i\epsilon_{\mu\nu\alpha\beta}\varepsilon_{D^{^{*}}_{_{q}}}^{\ast\mu}
\varepsilon_{D^{^{*}}}^{\ast\nu}p_{D^{^{*}}_{_{q}}}^{\alpha}p_{D^{^{*}}}^{\beta}
\frac{2V(m^2_{D^{^{*}}_{q}})}{m_{B}+m_{D^{^{*}}}}\right].
\end{eqnarray}

\subsection{Observables to be investigated}

We can get the total decay amplitudes in the RPV MSSM as
\begin{eqnarray}
\mathcal{M}(B\to D^{^{(*)}}D^{^{(*)}}_{_{q}})=\mathcal{M}^{\rm
SM}(B\to D^{^{(*)}}D^{^{(*)}}_{_{q}})+\mathcal{M}^{\rm RPV}(B\to
D^{^{(*)}}D^{^{(*)}}_{_{q}}). \label{ALLamp}
\end{eqnarray}
The branching ratio $\mathcal{B}$ reads as
\begin{eqnarray}
\mathcal{B}(B\to D^{^{(*)}}D^{^{(*)}}_{_{q}})=\frac{\tau_B |p_c
|}{8\pi m_B^2}\left|\mathcal{M}(B\to
D^{^{(*)}}D^{^{(*)}}_{_{q}})\right|^2,
\end{eqnarray}
where $\tau_B$ is the $B$ lifetime, $|p_c|$ is the center of mass
momentum  in the center of mass frame of $B$ meson. In $B\to
D^*D^*_q$ decays, the two vector mesons have the same helicity,
therefore three different polarization states, one longitudinal and
two transverse, are possible.  We define the corresponding
amplitudes as $\mathcal{M}_{0,\pm}$ in the helicity basis and
$\mathcal{M}_{L,\parallel,\perp}$ in the transversity basis, which
are related by $\mathcal{M}_{L}=\mathcal{M}_{0}$ and
$\mathcal{M}_{\parallel,\perp}=\frac{\mathcal{M}_+\pm\mathcal{M}_-}{\sqrt{2}}$.
Then we have
\begin{eqnarray}
\left|\mathcal{M}(B\to
D^*D^*_q)\right|^2=|\mathcal{M}_0|^2+|\mathcal{M}_+|^2+|\mathcal{M}_-|^2
=|\mathcal{M}_L|^2+|\mathcal{M}_\parallel|^2+|\mathcal{M}_\perp|^2.
\end{eqnarray}
The longitudinal polarization fraction $f_L$ and transverse
polarization fraction $f_\perp$ are defined by
\begin{eqnarray}
f_{L,\perp}(B\to
D^*D^*_q)&=&\frac{\Gamma_{L,\perp}}{\Gamma}=\frac{|\mathcal{M}_{L,\perp}|^2}
{|\mathcal{M}_L|^2+|\mathcal{M}_\parallel|^2+|\mathcal{M}_\perp|^2}.
\end{eqnarray}

In charged $B$ meson decays, where mixing effects are absent, the
only possible source of CPAs is
\begin{eqnarray}
\mathcal{A}_{\rm CP}^{k,{\rm
dir}}=\frac{\left|\mathcal{M}_k(B^-\rightarrow
\overline{f})/\mathcal{M}_k(B^+\rightarrow
f)\right|^2-1}{\left|\mathcal{M}_k(B^-\rightarrow
\overline{f})/\mathcal{M}_k(B^+\rightarrow f)\right|^2+1},
\end{eqnarray}
and $k=L,\parallel,\perp$ for $B^-\to D^*D^*_q$ decays and $k=L$ for
$B^-_u\to DD_q,DD^*_q,D^*D_q$ decays.  Then for $B^-_u\to D^*D^*_q$
decays, we have
\begin{eqnarray}
\mathcal{A}_{\rm CP}^{+,{\rm dir}}(B\to
D^*D^*_q)&=&\frac{\mathcal{A}_{\rm CP}^{\parallel,{\rm
dir}}|\mathcal{M}_\parallel|^2+\mathcal{A}_{\rm CP}^{L,{\rm
dir}}|\mathcal{M}_L|^2}
{|\mathcal{M}_\parallel|^2+|\mathcal{M}_L|^2}.\label{Eq:APdir}
\end{eqnarray}

For CPAs of neutral $B_q$ meson decays, there is an additional
complication due to $B^0_q-\bar{B}^0_q$ mixing. There are four cases
that one encounters for neutral $B_q$ decays, as discussed in Refs.
\cite{Gronau:1989zb,Soto:1988hf,Palmer:1994ec,Ali:1998gb}.
\begin{itemize}
\item[(i)] $B^0_q\to f, \bar{B}^0_q\to \bar{f}$, where $f$ or $\bar{f}$ is not
a common final state of $B^0_q$ and $\bar{B}^0_q$, for example
$B^0_q\to D^+D^-_s$.
\item[(ii)]
$B^0_q\to (f=\bar{f})\leftarrow\bar{B}^0_q$ with $f^{\rm CP}=\pm f$,
involving final states which are $CP$ eigenstates, $i.e.$, decays
such as $B^0_d\to D^+D^-,B^0_s\to D^+_sD^-_s$.
\item[(iii)]
$B^0_q\to (f=\bar{f})\leftarrow\bar{B}^0_q$ with $f^{\rm CP}\neq\pm
f$, involving final states which are not $CP$ eigenstates. They
include decays such as $B^0_q\to (VV)^0$, as the $VV$ states are not
$CP$ eigenstates.
\item[(iv)]
 $B^0_q\to (f\&\bar{f})\leftarrow \bar{B}^0_q$ with $f^{\rm CP}\neq
f$, $i.e.$, both $f$ and $\bar{f}$ are common final states of
$B^0_q$ and $\bar{B}^0_q$, but they are not $CP$ eigenstates. Decays
$B^0_d(\bar{B}^0_d)\to D^{*-}D^+,D^{-}D^{*+}$ and
$B^0_s(\bar{B}^0_s)\to D^{*-}_sD^+_s,D^{-}_sD^{*+}_s$  belong to
this case.
\end{itemize}

CPAs of neutral $B$ decays in case (i) are similar to CPAs of the
charged $B$ decays, and there are only direct CPAs $\mathcal{A}_{\rm
CP}^{\rm dir}$ since no mixing is involved for these decays. For
cases (ii) and (iii), their CPAs would involve $B^0_q-\bar{B}^0_q$
mixing. The time-dependent asymmetries can be conveniently expressed
as
\begin{eqnarray}
&&\mathcal{A}^k_{f}(t)=\mathcal{S}^k_f\sin(\Delta mt)-\mathcal{C}^k_f\cos(\Delta mt),\\
&&\mathcal{S}^k_f\equiv\frac{2\mbox{Im}(\lambda_k)}{1+\left|\lambda_k\right|^2},~~~~~~~~
\mathcal{C}^k_f\equiv\frac{1-\left|\lambda_k\right|^2}{1+\left|\lambda_k\right|^2},
\end{eqnarray}
where
$\lambda_k=\frac{q}{p}\frac{\mathcal{M}_k(\overline{B}^0\rightarrow
f)}{\mathcal{M}_k(B^0\rightarrow f)}$. In addition,
$\mathcal{S}^+_f$ and $\mathcal{C}^+_f$ can be obtained from the
similar relation given in Eq. (\ref{Eq:APdir}).

Case (iv) also involves mixing but requires additional formulae.
Here one studies the four time-dependent decay widths for
$B^0_q(t)\to f$, $\bar{B}^0_q(t)\to \bar{f}$, $B^0_q(t)\to \bar{f}$
and $\bar{B}^0_q(t)\to f$
\cite{Gronau:1989zb,Soto:1988hf,Palmer:1994ec,Ali:1998gb}. These
time-dependent widths can be expressed by four basic matrix elements
\cite{Palmer:1994ec}
\begin{eqnarray} g&=&\langle
f|\mathcal{H}_{eff}|B^0_q\rangle,~~~~h=\langle
f|\mathcal{H}_{eff}|\bar{B}^0_q\rangle, \nonumber\\
\bar{g}&=&\langle
\bar{f}|\mathcal{H}_{eff}|\bar{B}^0_q\rangle,~~~\bar{h}=\langle
\bar{f}|\mathcal{H}_{eff}|B^0_q\rangle,
\end{eqnarray}
which determine the decay matrix elements of $B^0_q\to f,\bar{f}$
and of $\bar{B}^0_q\to f,\bar{f}$ at $t=0$. We will study the
following quantities
\begin{eqnarray}
&&\mathcal{S}^k_f=\frac{2\mbox{Im}(\lambda'_k)}{1+\left|\lambda'_k\right|^2},~~~~
\mathcal{C}^k_f=\frac{1-\left|\lambda'_k\right|^2}{1+\left|\lambda'_k\right|^2},\\
&&
\mathcal{S}^k_{\bar{f}}=\frac{2\mbox{Im}(\lambda''_k)}{1+\left|\lambda''_k\right|^2},~~~~
\mathcal{C}^k_{\bar{f}}=\frac{1-\left|\lambda''_k\right|^2}{1+\left|\lambda''_k\right|^2},
\end{eqnarray}
with $\lambda'_k=(q/p)(h/g)$ and
$\lambda''_k=(q/p)(\bar{g}/\bar{h})$. The signatures of $CP$
violation are $\Gamma(\bar{B}^0_q(t)\to \bar{f}) \neq
\Gamma(B^0_q(t)\to f)$ and $\Gamma(\bar{B}^0_q(t)\to f) \neq
\Gamma(B^0_q(t)\to \bar{f})$, which means that
$\mathcal{C}_f\neq-\mathcal{C}_{\bar{f}}$ and/or
$\mathcal{S}_f\neq-\mathcal{S}_{\bar{f}}$.

\subsection{Input parameters}

Theoretical  input parameters are collected in Table
\ref{Table:inputs}. In our numerical results, we will use the input
parameters which are varied randomly within $1\sigma$ range.
\begin{table}[htbp]
\caption{Summary of theoretical input parameters and $\pm1 \sigma$
error ranges for sensitive parameters used in our numerical
calculations.}\vspace{0.3cm}
\begin{center}
\begin{tabular}{lr}\hline\hline
$m_{_{B_u}}=5.279~{\rm GeV},~m_{_{B_d}}=5.280~{\rm GeV},~m_{_{B_s}}=5.366~{\rm GeV},$$~M_{D^{0}}=1.865$ GeV, \\
$M_{D^{+}}=1.870$ GeV, $M_{D^{+}_s}=1.969$ GeV,$~M_{D^{*0}}=2.007$ GeV, $M_{D^{*+}}=2.010$ GeV, \\
$M_{D^{*+}_s}=2.107$
GeV,$~\overline{m}_b(\overline{m}_b)=(4.20\pm0.07)~{\rm
GeV},~\overline{m}_c(\overline{m}_c)=(1.25\pm0.09)~{\rm GeV},$\\
$\overline{m}_s(2~{\rm GeV})=(0.095\pm0.025)~{\rm GeV},$
$\overline{m}_u(2~{\rm GeV})=(0.0015\sim 0.0030)~{\rm GeV},$\\$\overline{m}_d(2~{\rm GeV})=(0.003\sim 0.007)~{\rm GeV},$ &\\
$\tau_{_{B_u}}=(1.638\pm0.011)~ps,~\tau_{_{B_d}}=(1.530\pm0.009)~ps,~\tau_{_{B_s}}=(1.425^{+0.041}_{-0.041})~ps.$&\cite{PDG}\\\hline
$|V_{ud}|=0.97430\pm0.00019,~|V_{us}|=0.22521^{+0.00083}_{-0.00082},~|V_{ub}|=0.00344^{+0.00022}_{-0.00017},$&\\
$|V_{cd}|=0.22508^{+0.00084}_{-0.00082},~|V_{cs}|=0.97350^{+0.00021}_{-0.00022},~|V_{cb}|=0.04045^{+0.00106}_{-0.00078},$&\\
$|V_{td}|=0.00841^{+0.00035}_{-0.00092},~|V_{ts}|=0.03972^{+0.00115}_{-0.00077},~|V_{tb}|=0.999176^{+0.000031}_{-0.000044},$&\\
$\alpha=\left(90.7^{+4.5}_{-2.9}\right)^\circ,~$$\beta=\left(21.7^{+1.0}_{-0.9}\right)^\circ,~$
$\gamma=\left(67.6^{+2.8}_{-4.5}\right)^\circ.$&\cite{CKMfit}\\\hline
$f_D=(0.201\pm0.003\pm0.017)$ GeV,~$f_{D_s}=(0.249\pm0.003\pm0.016)$
GeV. &\cite{Aubin:2005ar}
\\\hline\hline
\end{tabular}
\end{center}\label{Table:inputs}
\end{table}

We have several remarks on the input parameters:
\begin{itemize}
\item \underline{CKM matrix elements}: The weak phase $\gamma$ is
well constrained in the SM,  however, with the presence of R-parity
violation, this constraint may be relaxed. We will not take $\gamma$
within the SM range, but vary it randomly in the range of 0 to $\pi$
to obtain conservative limits on RPV couplings.

\item \underline{Decay constants}:  The decay constants of $D^*_q$ mesons have not been
directly measured in experiments so far. In the heavy-quark limit
$(m_c \to\infty )$, spin symmetry predicts that
$f_{D^*_{q}}=f_{D_{q}}$, and most theoretical predictions indicate
that symmetry-breaking corrections enhance the ratio
$f_{D^*_{q}}/f_{D_{q}}$ by $10\% - 20\%$
\cite{Neubert:1993mb,Neubert:1996qg}. Hence, we take
$f_{D^*_{q}}=(1.1 - 1.2)f_{D_{q}}$ as our input values.

\item \underline{Distribution amplitudes}: The distribution amplitudes of $D^{(*)}_q$ mesons are less constrained, and we
use the shape parameter $a_{D^{(*)}}=0.7\pm0.2$ and
$a_{D^{(*)}_s}=0.3\pm0.2$.

\item \underline{Form factors}: For the form factors involving $B\to D^{(*)}$ transitions, we
take expressions which include  perturbative QCD corrections induced
by hard gluon vertex corrections of $b\to c$ transitions and power
corrections in orders of $1/m_{b,c}$
\cite{Neubert:1991xw,Neubert:1992tg}. As for Isgur-Wise function
$\xi(\omega)$, we use the fit result
$\xi(\omega)=1-1.22(\omega-1)+0.85(\omega-1)^2$  from Ref.
\cite{Cheng:2003sm}.

\item \underline{Wilson coefficients}: We obtain Wilson coefficients
in terms of the expressions in \cite{Buchalla:1995vs}.

\item  \underline{RPV couplings}:  When we study the RPV effects, we  consider
only one RPV coupling product contributes at one time, neglecting
the interferences between different RPV coupling products, but
keeping their interferences  with the SM amplitude. We assume the
masses of sfermion are 100 GeV. For other values of the sfermion
masses, the bounds on the couplings in this paper can be easily
obtained by scaling them by factor
$\tilde{f}^2\equiv(\frac{m_{\tilde{f}}}{100~\rm{GeV}})^2$.
\end{itemize}

\section{Numerical results and discussions}

In this section we summarize  our numerical results and analysis in
the exclusive color-allowed $b\to c\bar{c} q$ decays. First, we will
show our estimates in the SM with  full theoretical uncertainties of
sensitive parameters. Then, we will investigate the RPV effects in
the decays. We will constrain relevant RPV couplings only from quite
highly consistent experimental data and show the RPV MSSM
predictions for the other observables, which have not been measured
yet or have less consistency among different collaborations.

\subsection{Exclusive color-allowed  $b\to c\bar{c} d$ decays }

Decays $\bar{B}^0_{d}\to D^{^{(*)+}}D^{^{(*)-}}$, $B^-_{u}\to
D^{^{(*)0}}D^{^{(*)-}}$ and $\bar{B}^0_{s}\to
D^{^{(*)+}}_sD^{^{(*)-}}$ are dominated by the color-allowed $b\to
c\bar{c}d$ tree diagram, but involve small penguin pollution from
the $b\to u\bar{u}d$ transition carrying a different weak phase.
These decays involve the same set of  RPV coupling constants
$\lambda''^*_{232}\lambda''_{212}$ and
$\lambda'^*_{i23}\lambda'_{i21}$ at tree level due to squark and
slepton exchanges, respectively.  For $\bar{B}^0_{d}\to
D^{^{(*)+}}D^{^{(*)-}}$ and $B^-_{u}\to D^{^{(*)0}}D^{^{(*)-}}$
processes, a few observables have been measured by B{\footnotesize
A}B{\footnotesize AR} and Belle collaborations. The latest
experimental data and their weight averages are summarized in Table
\ref{Table:btoccd data}. We can see almost all physical quantities
have been consistently measured between B{\footnotesize
A}B{\footnotesize AR} and Belle, and only
$\mathcal{B}(\bar{B}^0_d\to D^{+}D^-,D^{*\pm}D^\mp)$,
$\mathcal{C}(\bar{B}^0_d\to D^+D^-)$  and
$\mathcal{C}(B^0_d,\bar{B}^0_d\to D^+D^{*-})$ have low consistency
between B{\footnotesize A}B{\footnotesize AR} and Belle.

\begin{table}[t]
\caption{ Experimental data for $\bar{B}^0_{d}\to
D^{^{(*)+}}D^{^{(*)-}}$ and $B^-_{u}\to D^{^{(*)0}}D^{^{(*)-}}$
decays from B{\footnotesize A}B{\footnotesize AR} and Belle. The
branching ratios $(\mathcal{B})$ are in units of $10^{-4}$. The
scale factor $S$ is defined in introduction part of Ref. \cite{PDG},
and $S>1$ often indicates that the measurements are inconsistent. }
\begin{center}{\footnotesize
\begin{tabular}{lrrcr}\hline\hline
~~~Observable &B{\footnotesize A}B{\footnotesize AR}
~~~~~~~~~~&Belle~~~~~~~~~~~~&Average& $S$~~\\\hline
$\mathcal{B}(\bar{B}^0_d\to D^+D^-)$&$2.8\pm0.4\pm0.5$
\cite{Aubert:2006ia} &$1.97\pm0.20\pm0.20$
\cite{Fratina:2007zk}&$2.1\pm0.3$&$\bf{1.2}$\\\hline
$\mathcal{B}(\bar{B}^0_d\to D^{*\pm}D^\mp)$&$5.7\pm0.7\pm0.7$
 \cite{Aubert:2006ia}&$11.7\pm2.6^{+2.2}_{-2.5}$ \cite{Abe:2002zy}&$6.1\pm1.5$&$\bf{1.6}$\\\hline
$\mathcal{B}(\bar{B}^0_d\to D^{*+}D^{*-})$&$8.1\pm0.6\pm1.0$
 \cite{Aubert:2006ia}&$8.1\pm0.8\pm1.1$ \cite{Miyake:2005qb}&$8.1\pm0.9$&$\leq1.0$\\\hline
$\mathcal{B}(B^-_u\to D^{0}D^{-})$&$3.8\pm0.6\pm0.5$
\cite{Aubert:2006ia}&$3.85\pm0.31\pm0.38$
\cite{Abe:2007sk}&$3.8\pm0.4$&$\leq1.0$\\\hline
$\mathcal{B}(B^-_u\to D^{*0}D^{-})$&$6.3\pm1.4\pm1.0$
\cite{Aubert:2006ia}&&&\\\hline
$\mathcal{B}(B^-_u\to
D^{0}D^{*-})$&$3.6\pm0.5\pm0.4$
\cite{Aubert:2006ia}&$4.57\pm0.71\pm0.56$
\cite{Majumder:2005gy}&$3.9\pm0.5$&$\leq1.0$\\\hline
$\mathcal{B}(B^-_u\to D^{*0}D^{*-})$&$8.1\pm1.2\pm1.2$
\cite{Aubert:2006ia}&&&\\\hline
$\mathcal{C}(\bar{B}^0_d\to D^+D^-)$&$-0.07\pm0.23\pm0.03$
\cite{:2008aw}&$-0.91\pm0.23\pm0.06$
\cite{Fratina:2007zk}&$-0.48\pm0.42$&$\bf{2.5}$\\\hline
$\mathcal{C}(B^0_d,\bar{B}^0_d\to D^{*+}D^-)$&$0.08\pm0.17\pm0.04$
\cite{:2008aw}&$-0.37\pm0.22\pm0.06$
\cite{Aushev:2004uc}&$-0.09\pm0.22$&$\bf{1.6}$\\\hline
$\mathcal{C}(B^0_d,\bar{B}^0_d\to D^+D^{*-})$&$0.00\pm0.17\pm0.03$
\cite{:2008aw}&$0.23\pm0.25\pm0.06$
\cite{Aushev:2004uc}&$0.07\pm0.14$&$\leq1.0$\\\hline
$\mathcal{C}^+(\bar{B}^0_d\to D^{*+}D^{*-})$&$0.00\pm0.12\pm0.02$
\cite{:2008aw}&$-0.15\pm0.13\pm0.04$
\cite{Vervink:2008dv}&$-0.07\pm0.09$&$\leq1.0$\\\hline
$\mathcal{A}^{\rm dir}_{\rm CP}(B^-_u\to
D^0D^-)$&$-0.13\pm0.14\pm0.02$
\cite{Aubert:2006ia}&$0.00\pm0.08\pm0.02$
\cite{Abe:2007sk}&$-0.03\pm0.07$&$\leq1.0$\\\hline $\mathcal{A}^{\rm
dir}_{\rm CP}(B^-_u\to D^{*0}D^{-})$&$0.13\pm0.18\pm0.04$
\cite{Aubert:2006ia}&&&\\\hline
 $\mathcal{A}^{\rm dir}_{\rm CP}(B^-_u\to
D^{0}D^{*-})$&$-0.06\pm0.13\pm0.02$
\cite{Aubert:2006ia}&$0.15\pm0.15\pm0.05$
\cite{Majumder:2005gy}&$0.03\pm0.10$&$\leq1.0$\\\hline
$\mathcal{A}^{+, \rm dir}_{\rm CP}(B^-_u\to
D^{*0}D^{*-})$&$-0.15\pm0.11\pm0.02$ \cite{Aubert:2006ia}&&&\\\hline
$\mathcal{S}(\bar{B}^0_d\to D^+D^-)$&$-0.63\pm0.36\pm0.05$
\cite{:2008aw}&$-1.13\pm0.37\pm0.09$
\cite{Fratina:2007zk}&$-0.87\pm0.26$&$\leq1.0$\\\hline
$\mathcal{S}(B^0_d,\bar{B}^0_d\to D^{*+}D^-)$&$-0.62\pm0.21\pm0.03$
\cite{:2008aw}&$-0.55\pm0.39\pm0.12$
\cite{Aushev:2004uc}&$-0.61\pm0.19$&$\leq1.0$\\\hline
$\mathcal{S}(B^0_d,\bar{B}^0_d\to D^+D^{*-})$&$-0.73\pm0.23\pm0.05$
\cite{:2008aw}&$-0.96\pm0.43\pm0.12$
\cite{Aushev:2004uc}&$-0.78\pm0.21$&$\leq1.0$\\\hline
$\mathcal{S}^+(\bar{B}^0_d\to D^{*+}D^{*-})$&$-0.76\pm0.16\pm0.04$
\cite{:2008aw}&$-0.96\pm0.25^{+0.12}_{-0.16}$
\cite{Vervink:2008dv}&$-0.81\pm0.14$&$\leq1.0$\\\hline
$f_\perp(\bar{B}^0_d\to D^{*+}D^{*-})$&$0.158\pm0.028\pm0.006$
\cite{:2008aw}&$0.125\pm0.043\pm0.023$
\cite{Vervink:2008dv}&$0.150\pm0.025$&$\leq1.0$\\\hline
$f_L(\bar{B}^0_d\to D^{*+}D^{*-})$&&$0.57\pm0.08\pm0.02$
\cite{Miyake:2005qb}&&\\\hline \hline
\end{tabular}}
\end{center}\label{Table:btoccd data}
\end{table}

Our SM estimates predicted within the theoretical uncertainties of
input parameters are given in the second columns of Table
\ref{Table:btoccdBRFL} and Table \ref{Table:btoccdACP}. Theoretical
predictions for the branching ratios and the polarization fractions
are given in Table \ref{Table:btoccdBRFL}. CPA predictions are given
in Table \ref{Table:btoccdACP}.
All the branching ratios are above $10^{-4}$ order. The direct CPAs
are expected to be quite small. All mixing-induced CPAs  of
$\bar{B}^0_d$ decays are very large (about $-0.7$).  There is  an
obvious signature of the mixing-induced $CP$ violations in
$\bar{B}^0_s\to D^{*+}_sD^{-}_s,D^{+}_sD^{*-}_s$ decays since
$\mathcal{S}(B^0_s,\bar{B}^0_s\to D^{*+}_sD^{-}_s)\neq
-\mathcal{S}(B^0_s,\bar{B}^0_s\to D^{+}_sD^{*-}_s)$, which are
consistent with the experimental measurements. In addition, for
$\bar{B}^0_{d}\to D^{^{*+}}D^{^{*-}}$, $B^-_{u}\to
D^{^{*0}}D^{^{*-}}$ and $\bar{B}^0_{s}\to D^{^{*+}}_sD^{^{*-}}$
decays, the longitudinal and transverse polarization fractions can
be precisely predicted, and are about $\sim$0.5 and $\sim$0.1,
respectively.
 Comparing present experimental data in Table
\ref{Table:btoccd data} with the SM predictions in Table
\ref{Table:btoccdBRFL} and Table \ref{Table:btoccdACP}, we can find that
all measured quantities
 agree with the SM expectations within the error ranges except
 $\mathcal{C}(B^0_d,\bar{B}^0_d\to D^+D^-)$ from Belle.

\begin{table}[tb]
\caption{\small Theoretical predictions for CP-averaged branching
ratios (in units of $10^{-4}$) and ratios of polarization (in units
of $10^{-2}$) in exclusive color-allowed $b\to c\bar{c}d$ decays.
The second column gives the SM predictions with the theoretical
uncertainties of input parameters. The last two columns are the RPV
MSSM predictions with different RPV couplings considering the input
parameter uncertainties and experimental errors.  }
\begin{center}
\begin{tabular}{lccc}\hline\hline
~~~~Observable&SM &MSSM w/$~~\lambda''^*_{232}\lambda''_{212}$& MSSM
w/~~$\lambda'^*_{i23}\lambda'_{i21}$\\\hline
$\mathcal{B}(\bar{B}^0_d\to
D^+D^-)$&$[2.35,4.15]$&$[2.77,3.80]$&$[2.77,4.39]$\\\hline
$\mathcal{B}(\bar{B}^0_d\to
D^{*+}D^-)$&$[2.27,3.96]$&$[2.87,4.22]$&$[2.30,4.59]$\\\hline
$\mathcal{B}(\bar{B}^0_d\to
D^+D^{*-})$&$[2.56,5.04]$&$[3.31,4.65]$&\\\hline
 $\mathcal{B}(\bar{B}^0_d\to D^{*\pm}D^\mp)$&$[4.84,8.95]$&$[6.18,8.70]$&$[5.21,8.84]$\\\hline
$\mathcal{B}(\bar{B}^0_d\to
D^{*+}D^{*-})$&$[6.21,12.22]$&$[7.05,8.90]$&\\\hline
$\mathcal{B}(B^-_u\to
D^{0}D^{-})$&$[2.53,4.43]$&$[3.00,4.04]$&$[3.00,4.68]$\\\hline
$\mathcal{B}(B^-_u\to
D^{*0}D^{-})$&$[2.42,4.27]$&$[3.07,4.53]$&$[2.25,4.75]$\\\hline
$\mathcal{B}(B^-_u\to
D^{0}D^{*-})$&$[2.73,5.42]$&$[3.55,4.96]$&\\\hline
$\mathcal{B}(B^-_u\to
D^{*0}D^{*-})$&$[6.61,13.10]$&$[7.54,10.67]$&\\\hline
$\mathcal{B}(\bar{B}^0_s\to
D^+_sD^-)$&$[2.33,4.20]$&$[2.76,3.81]$&$[2.77,4.48]$\\\hline
$\mathcal{B}(\bar{B}^0_s\to
D^{*+}_sD^-)$&$[2.24,4.00]$&$[2.81,4.25]$&$[2.08,4.42]$\\\hline
$\mathcal{B}(\bar{B}^0_s\to
D^+_sD^{*-})$&$[2.54,5.03]$&$[3.27,4.69]$&\\\hline
$\mathcal{B}(\bar{B}^0_s\to
D^{*+}_sD^{*-})$&$[6.15,12.10]$&$[6.92,10.04]$&\\\hline
$f_L(\bar{B}^0_d\to
D^{*+}D^{*-})$&$[52.40,52.97]$&$[50.35,52.53]$&\\\hline
$f_L(B^-_u\to
D^{*0}D^{*-})$&$[52.43,53.02]$&$[50.37,52.56]$&\\\hline
$f_L(\bar{B}^0_s\to
D^{*+}_sD^{*-})$&$[52.56,53.16]$&$[50.60,52.70]$&\\\hline
$f_\perp(\bar{B}^0_d\to
D^{*+}D^{*-})$&$[8.82,9.51]$&$[10.00,12.94]$&\\\hline
$f_\perp(B^-_u\to
D^{*0}D^{*-})$&$[8.84,9.53]$&$[10.03,12.97]$&\\\hline
$f_\perp(\bar{B}^0_s\to
D^{*+}_sD^{*-})$&$[8.35,9.05]$&$[9.50,12.34]$&\\\hline \hline
\end{tabular}
\end{center}\label{Table:btoccdBRFL}
\end{table}

\begin{table}[tb]
\caption{\small Theoretical predictions for CPAs (in units of
$10^{-2}$) in exclusive color-allowed $b\to c\bar{c}d$ decays. }
\begin{center}
\begin{tabular}{lccc}\hline\hline
~~~~~~~Observable& SM &MSSM w/ $\lambda''^*_{232}\lambda''_{212}$&
MSSM w/ $\lambda'^*_{i23}\lambda'_{i21}$
\\\hline
$\mathcal{S}(B^0_d,\bar{B}^0_d\to
D^+D^-)$&$[-78.00,-71.67]$&$[-97.52,-52.66]$&$[-99.83,-35.16]$\\\hline
$\mathcal{S}(B^0_d,\bar{B}^0_d\to
D^{*+}D^-)$&$[-70.40,-64.55]$&$[-81.77,-32.17]$&$[-98.01,-55.02]$\\\hline
$\mathcal{S}(B^0_d,\bar{B}^0_d\to
D^+D^{*-})$&$[-72.17,-66.83]$&$[-83.29,-36.15]$&$[-98.30,-57.69]$\\\hline
$\mathcal{S}^+(B^0_d,\bar{B}^0_d\to
D^{*+}D^{*-})$&$[-72.73,-67.77]$&$[-95.18,-53.01]$&\\\hline
$\mathcal{C}(B^0_d,\bar{B}^0_d\to
D^+D^-)$&$[-6.03,-3.87]$&$[-7.61,0.92]$&$[-11.05,2.59]$\\\hline
$\mathcal{C}(B^0_d,\bar{B}^0_d\to
D^{*+}D^-)$&$[3.36,13.83]$&$[1.38,14.75]$&$[-13.35,21.30]$\\\hline
$\mathcal{C}(B^0_d,\bar{B}^0_d\to
D^+D^{*-})$&$[-14.44,-3.53]$&$[-14.83,-1.45]$&$[-21.00,13.28]$\\\hline
$\mathcal{C}^+(B^0_d,\bar{B}^0_d\to
D^{*+}D^{*-})$&$[-1.35,-1.03]$&$[-1.83,0.20]$&\\\hline
$\mathcal{A}^{\rm dir}_{\rm CP}(B^-_u\to
D^0D^-)$&$[3.87,6.03]$&$[-0.92,7.61]$&$[-2.59,11.05]$\\\hline
$\mathcal{A}^{\rm dir}_{\rm CP}(B^-_u\to
D^{*0}D^-)$&$[-1.15,-0.45]$&$[-1.10,0.29]$&$[-1.99,0.23]$\\\hline
$\mathcal{A}^{\rm dir}_{\rm CP}(B^-_u\to
D^0D^{*-})$&$[1.03,1.35]$&$[-0.40,1.42]$&\\\hline
$\mathcal{A}^{+,dir}_{CP}(B^-_u\to
D^{*0}D^{*-})$&$[1.03,1.35]$&$[-0.20,1.83]$&\\\hline
$\mathcal{A}^{\rm dir}_{\rm CP}(\bar{B}^0_s\to
D^+_sD^-)$&$[3.87,6.03]$&$[-0.92,7.61]$&$[-2.59,11.05]$\\\hline
$\mathcal{A}^{\rm dir}_{\rm CP}(\bar{B}^0_s\to
D^{*+}_sD^-)$&$[-1.15,-0.45]$&$[-1.10,0.29]$&$[-1.99,0.23]$\\\hline
$\mathcal{A}^{\rm dir}_{\rm CP}(\bar{B}^0_s\to
D^+_sD^{*-})$&$[1.03,1.35]$&$[-0.40,1.42]$&\\\hline
$\mathcal{A}^{+,{\rm dir}}_{\rm CP}(\bar{B}^0_s\to
D^{*+}_sD^{*-})$&$[1.03,1.35]$&$[-0.20,1.83]$&\\\hline \hline
\end{tabular}
\end{center}\label{Table:btoccdACP}
\end{table}

\begin{figure}[b]
\begin{center}
\begin{tabular}{c}
\includegraphics[scale=1]{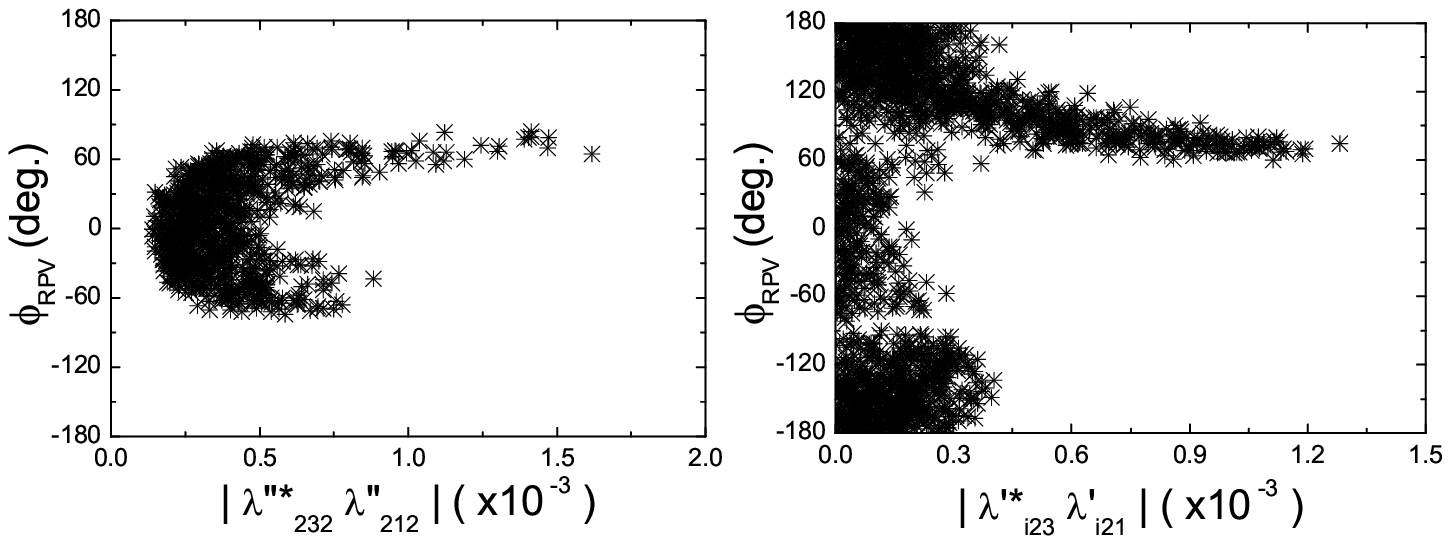}
\end{tabular}
\end{center}
\vspace{-0.6cm} \caption{\small Allowed parameter spaces for
relevant RPV couplings constrained by $\bar{B}^0_{d}\to
D^{^{(*)+}}D^{^{(*)-}}$ and $B^-_{u}\to D^{^{(*)0}}D^{^{(*)-}}$, where
$\phi_{\rm RPV}$ denotes the RPV weak phase.} \label{btoccdbounds}
\end{figure}

We now turn to explore the  RPV effects in $\bar{B}^0_{d}\to
D^{^{(*)+}}D^{^{(*)-}}$, $B^-_{u}\to D^{^{(*)0}}D^{^{(*)-}}$ and
$\bar{B}^0_{s}\to D^{^{(*)+}}_sD^{^{(*)-}}$ decays. The most
conservative existing experimental bounds are used  in our analysis. We
choose the averaged data, which have highly consistent measurements
between B{\footnotesize A}B{\footnotesize AR} and Belle (defined as a scale factor $S \leq
1$), and varied randomly within $2\sigma$ ranges to constrain the
RPV effects. The current experimental data and theoretical input
parameters are not yet precise enough to set absolute bounds on the
relative RPV couplings. We obtain the allowed scattering spaces of
the RPV couplings $\lambda''^*_{232}\lambda''_{212}$ and
$\lambda'^*_{i23}\lambda'_{i21}$ as displayed in Fig.
\ref{btoccdbounds}. These survived parameter spaces are not in
 conflict with the above mentioned highly consistent experimental data in
$\bar{B}^0_{d}\to D^{^{(*)+}}D^{^{(*)-}}$ and $B^-_{u}\to
D^{^{(*)0}}D^{^{(*)-}}$ decays.

The squark exchange coupling $\lambda''^*_{232}\lambda''_{212}$
contributes to all twelve $\bar{B}^0_{d}\to D^{^{(*)+}}D^{^{(*)-}}$,
$B^-_{u}\to D^{^{(*)0}}D^{^{(*)-}}$ and $\bar{B}^0_{s}\to
D^{^{(*)+}}_sD^{^{(*)-}}$   decay modes. The allowed space of
$\lambda''^*_{232}\lambda''_{212}$ is shown in the left plot of Fig.
\ref{btoccdbounds}. Its magnitude
$|\lambda''^*_{232}\lambda''_{212}|$ and its RPV weak phase
$\phi_{\rm RPV}$ have been constrained significantly. We obtain
$|\lambda''^*_{232}\lambda''_{212}|\in [0.14, 1.62]\times 10^{-3}$
and $ \phi_{\rm RPV}\in [-75^\circ,84^\circ]$. The right plot of
Fig. \ref{btoccdbounds} displays the allowed space of the RPV
couplings $\lambda'^*_{i23}\lambda'_{i21}$ due to slepton exchanges,
which contributes only to six decay modes $\bar{B}^0_{d}\to
D^{^{(*)+}}D^{^{-}}$, $B^-_{u}\to D^{^{(*)0}}D^{^{-}}$ and
$\bar{B}^0_{s}\to D^{^{(*)+}}_sD^{^{-}}$. The magnitudes
$|\lambda'^*_{i23}\lambda'_{i21}|$ have been  limited within
$|\lambda'^*_{i23}\lambda'_{i21}|\leq1.28\times 10^{-3}$, and the
corresponding RPV weak phase $ \phi_{\rm RPV}$ for the range
$|\lambda'^*_{i23}\lambda'_{i21}|\leq0.4\times10^{-3}$ is not
constrained so much, however, the RPV weak phase for
$|\lambda'^*_{i23}\lambda'_{i21}|\in[0.4,1.3] \times 10^{-3}$ is
very narrow.

Using the constrained  parameter spaces shown in Fig.
\ref{btoccdbounds}, one may  predict the RPV effects on the other
quantities which have not been measured yet or have less consistent
measurements  between B{\footnotesize A}B{\footnotesize AR} and
Belle.  With the expressions for $\mathcal{B}$,
$\mathcal{C},~\mathcal{S},~\mathcal{A}^{\rm dir}_{\rm CP}$, $f_{L}$
and $f_{\perp}$ at hand, we perform a scan on the input parameters
and the constrained RPV couplings. Then we obtain the RPV MSSM
predictions with different RPV coupling, whose numerical results are
summarized in the last two columns of Table \ref{Table:btoccdBRFL}
and Table \ref{Table:btoccdACP}.

The contributions of $\lambda''^*_{232}\lambda''_{212}$ due to
squark exchange are summarized  in the third columns of Table
\ref{Table:btoccdBRFL} and Table \ref{Table:btoccdACP}. In Table
\ref{Table:btoccdBRFL}, comparing with the SM predictions, we find
$\lambda''^*_{232}\lambda''_{212}$ coupling could not affect all
branching ratios much. Three $f_L(B\to D^{*}D^{*})$ and three
$f_\perp(B\to D^{*}D^{*})$ are slightly decreased and increased by
$\lambda''^*_{232}\lambda''_{212}$ coupling, respectively, and their
allowed ranges are scarcely magnified by this coupling. As given in
Table \ref{Table:btoccdACP}, the $\lambda''^*_{232}\lambda''_{212}$
contributions could greatly enlarge the ranges of four
$\mathcal{S}(B^0_d,\bar{B}^0_{d}\to D^{^{(*)+}}D^{^{(*)-}})$. The
effects of $\lambda''^*_{232}\lambda''_{212}$ coupling could extend
a little bit the allowed regions of four
$\mathcal{C}(B^0_d,\bar{B}^0_d\to D^{(*)+}D^{(*)-})$ and eight
$\mathcal{A}^{\rm dir}_{\rm CP}(B^-_{u}\to
D^{(*)0}D^{(*)-},\bar{B}^0_{s}\to D^{(*)+}_sD^{(*)-})$, too. But
this squark exchange coupling cannot explain the large
$\mathcal{C}(B^0_d,\bar{B}^0_{d}\to D^{^{+}}D^{^{-}})$ from Belle.
The predictions including  slepton exchange couplings
$\lambda'^*_{i23}\lambda'_{i21}$ are listed  in the last columns of
Table \ref{Table:btoccdBRFL} and Table \ref{Table:btoccdACP}.
 The
$\lambda'^*_{i23}\lambda'_{i21}$ couplings do not give  very big
effects on the relevant branching ratios, but  could significantly
magnify  the ranges of $\mathcal{S}(B^0_d,\bar{B}^0_{d}\to
D^{^{(*)+}}D^{^{-}})$ from their SM predictions as well as extend
the ranges of $\mathcal{C}(B^0_d,\bar{B}^0_d\to D^{(*)+}D^{-})$ and
$\mathcal{A}^{\rm dir}_{\rm CP}(B^-_{u}\to
D^{(*)0}D^{-},\bar{B}^0_{s}\to D^{(*)+}_sD^{-})$. The lower limits
of $\mathcal{C}(B^0_d,\bar{B}^0_d\to D^{(*)+}D^{-})$ could  be
reduced by these slepton exchange couplings, too,  but slepton
exchange coupling effects are still not large enough to explain the
large $\mathcal{C}(B^0_d,\bar{B}^0_{d}\to D^{^{+}}D^{^{-}})$ from
Belle.

It is worth noting that our investigation of the color-allowed $b\to
c\bar{c}d$ decays was motivated by the large direct CPA of
$\bar{B}^0_{d}\to D^{^{+}}D^{^{-}}$ reported by Belle
\cite{Fratina:2007zk}, which has not been confirmed by
B{\footnotesize A}B{\footnotesize AR} and contradicted the SM
prediction. Relative RPV couplings, constrained by all consistent
measurements in $\bar{B}^0_{d}\to D^{^{(*)+}}D^{^{(*)-}}$ and
$B^-_{u}\to D^{^{(*)0}}D^{^{(*)-}}$ systems, could slightly enlarge
the range of $\mathcal{C}(B^0_d,\bar{B}^0_{d}\to D^{^{+}}D^{^{-}})$.
Our RPV MSSM prediction for this observable is coincident with the
B{\footnotesize A}B{\footnotesize AR} measurement, but still cannot
explain the Belle measurement. The unparticle interaction  has
positive effects on $\mathcal{C}(B^0_d,\bar{B}^0_{d}\to
D^{^{+}}D^{^{-}})$ as obtained in Ref. \cite{Zwicky:2007vv}, in
which the author  used only experimental constraints of
$\mathcal{B}(\bar{B}^0_{d}\to D^{^{+}}D^{^{-}})$. Note also that
very large value of $\mathcal{C}(B^0_d,\bar{B}^0_{d}\to
D^{^{+}}D^{^{-}})$  could be obtained by
unparticle interaction, however, with the sign opposite to the Belle
measurement.

\begin{figure}[t]
\begin{center}
\begin{tabular}{c}
\includegraphics[scale=1]{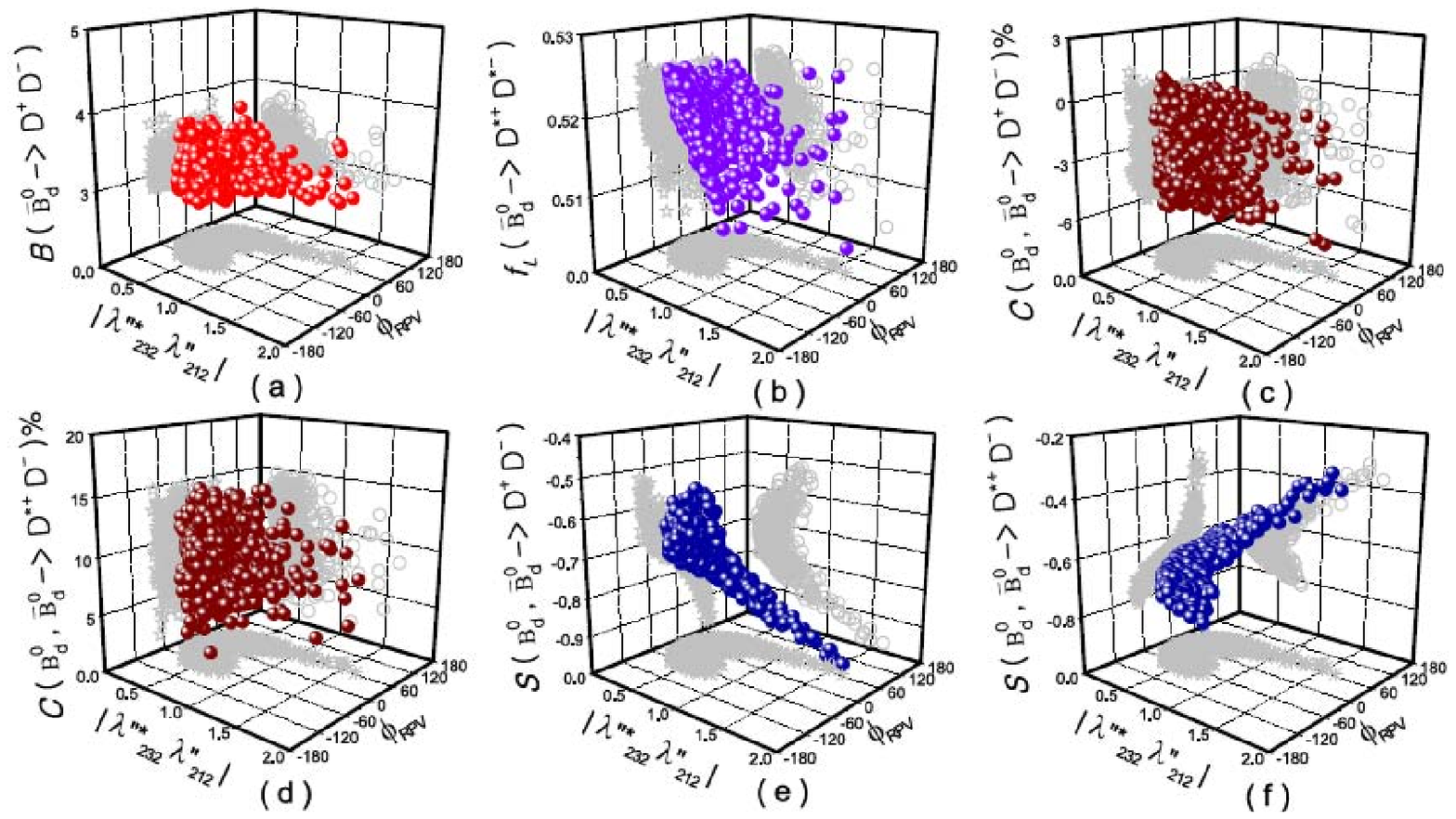}
\end{tabular}
\end{center}
\vspace{-0.6cm} \caption{\small Effects of RPV coupling
$\lambda''^*_{232}\lambda''_{212}$  in $\bar{B}^0_{d}\to
D^{^{(*)+}}D^{^{(*)-}}$ decays, where
$|\lambda''^*_{232}\lambda''_{212}|$ is in units of
$10^{-3}$, and $\mathcal{B}$ in units of $10^{-4}$.}\label{btoccdlpp}
\vspace{1.2cm}
\begin{center}
\begin{tabular}{c}
\includegraphics[scale=1]{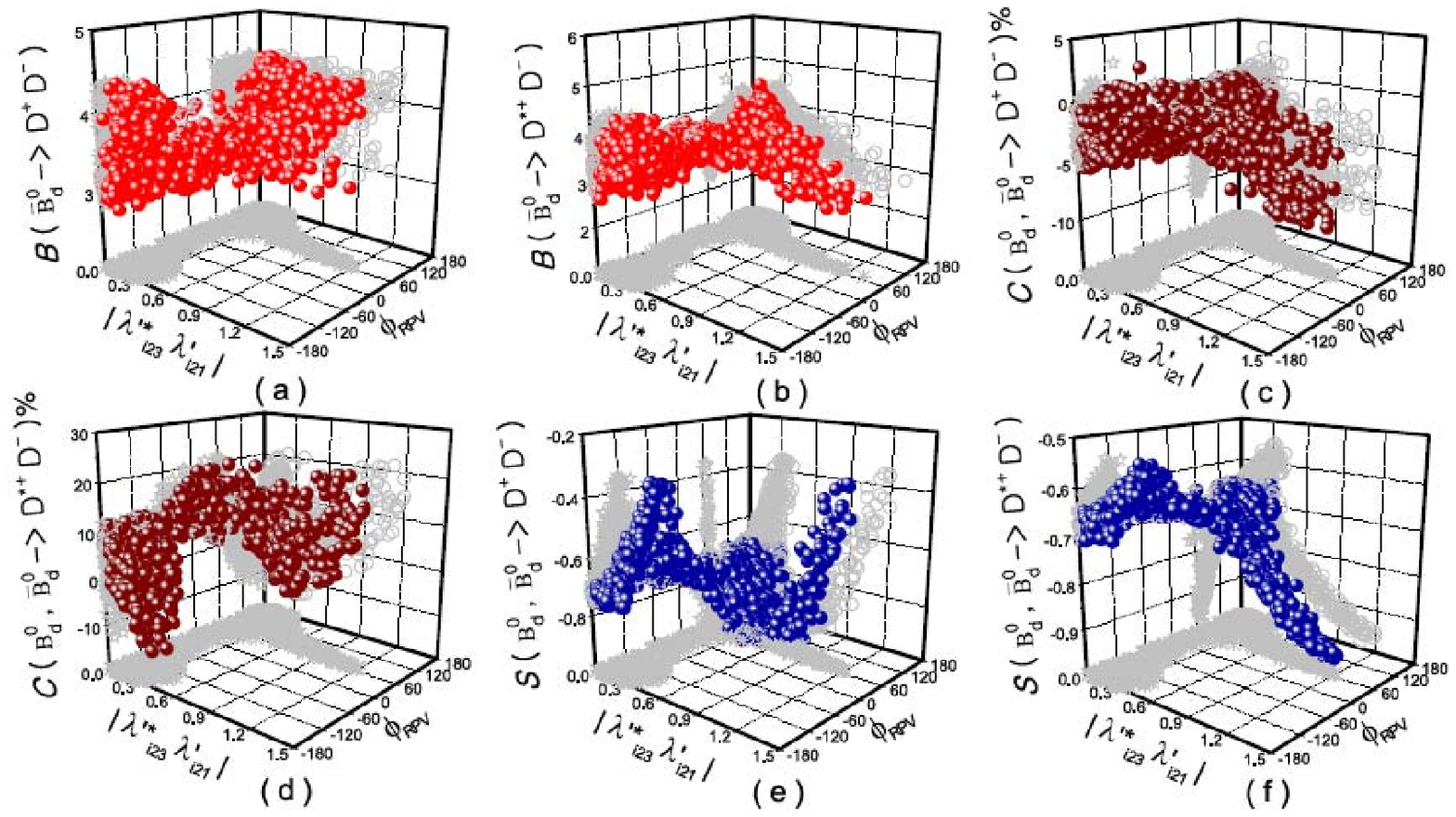}
\end{tabular}
\end{center}
\vspace{-0.6cm} \caption{\small Effects of RPV coupling
$\lambda'^*_{i23}\lambda'_{i21}$  in $\bar{B}^0_{d}\to
D^{^{(*)+}}D^{^{-}}$ decays, where  $|\lambda'^*_{i23}\lambda'_{i21}|$
is in units of $10^{-3}$, and $\mathcal{B}$ in units of $10^{-4}$. } \label{btoccdlp}
\end{figure}

For each RPV coupling product, we can present correlations of
physical quantities  within the constrained parameter spaces
displayed in Fig. \ref{btoccdbounds}  by the three-dimensional
scatter plots. RPV coupling   contributions to $\bar{B}^0_{d}\to
D^{^{(*)+}}D^{^{(*)-}}$, $B^-_{u}\to D^{^{(*)0}}D^{^{(*)-}}$ and
$\bar{B}^0_{s}\to D^{^{(*)+}}_sD^{^{(*)-}}$ decays are  very similar
to each other. So we will take an example for a few observables of
$\bar{B}^0_{d}\to D^{^{(*)+}}D^{^{(*)-}}$ decays to illustrate
RPV coupling effects. Effects of RPV couplings
 $\lambda''^*_{232}\lambda''_{212}$ and $\lambda'^*_{i23}\lambda'_{i21}$ on
observables of $\bar{B}^0_{d}\to
D^{^{(*)+}}D^{^{(*)-}}$ decays are shown in Fig. \ref{btoccdlpp} and
Fig. \ref{btoccdlp}, respectively.

In Fig. \ref{btoccdlpp},  we plot $\mathcal{B}$, $f_{L}$,
$\mathcal{C}$ and $\mathcal{S}$  as functions of
$\lambda''^*_{232}\lambda''_{212}$. Three-dimensional scatter plot
Fig. \ref{btoccdlpp} (a) shows $\mathcal{B}(\bar{B}^0_d\to
D^{+}D^{-})$ correlated with $|\lambda''^*_{232}\lambda''_{212}|$
and its phase $\phi_{\rm RPV}$. We also give projections to three
perpendicular
 planes, where the $|\lambda''^*_{232}\lambda''_{212}|$-$\phi_{\rm RPV}$ plane displays the
 constrained
 regions of $\lambda''^*_{232}\lambda''_{212}$, as the left plot of Fig. \ref{btoccdbounds}.
 It is shown that $\mathcal{B}(\bar{B}^0_d\to
D^{+}D^{-})$ is little decreasing with
$|\lambda''^*_{232}\lambda''_{212}|$  on the
$\mathcal{B}$-$|\lambda''^*_{232}\lambda''_{212}|$ plane. {}From the
$\mathcal{B}$-$\phi_{\rm RPV}$ plane,
 we see that $\mathcal{B}(\bar{B}^0_d\to
D^{+}D^{-})$ is not sensitive to $\phi_{\rm RPV}$. All other
branching ratios have similar trends with
$\lambda''^*_{232}\lambda''_{212}$ coupling. {}From Fig.
\ref{btoccdlpp} (b-d), we can see $f_L(\bar{B}^0_d\to D^{*+}D^{*-})$
and $\mathcal{C}(B^0_d,\bar{B}^0_d\to D^{+}D^{-},D^{*+}D^{-})$ are
not very sensitive with $|\lambda''^*_{232}\lambda''_{212}|$ and
$\phi_{\rm RPV}$. RPV coupling $\lambda''^*_{232}\lambda''_{212}$
contributions to $\mathcal{S}(B^0_d,\bar{B}^0_d\to
D^{+}D^{-},D^{*+}D^{*-})$ are also very similar to  each other. So
we  take an example for $\mathcal{S}(B^0_d,\bar{B}^0_d\to
D^{+}D^{-})$ shown in Fig. \ref{btoccdlpp} (e)  to illustrate the
RPV coupling effects. $\mathcal{S}(B^0_d,\bar{B}^0_{d}\to
D^{^{+}}D^{^{-}})$ is decreasing (increasing) with
$|\lambda''^*_{232}\lambda''_{212}|$ when $\phi_{\rm RPV}>0$
($\phi_{\rm RPV}<0$), and it is decreasing with $\phi_{\rm RPV}$.
$\mathcal{S}(B^0_d,\bar{B}^0_d\to D^{*+}D^{-},D^{+}D^{*-})$ have
totally different trends to $\mathcal{S}(B^0_d,\bar{B}^0_{d}\to
D^{^{+}}D^{^{-}})$ with $|\lambda''^*_{232}\lambda''_{212}|$ and
$\phi_{\rm RPV}$, and we show only the squark exchange effects on
$\mathcal{S}(B^0_d,\bar{B}^0_{d}\to D^{^{*+}}D^{^{-}})$ in Fig.
\ref{btoccdlpp} (f).

Fig. \ref{btoccdlp} gives the effects of the slepton exchange
couplings $\lambda'^*_{i23}\lambda'_{i21}$ in $\bar{B}^0_{d}\to
D^{^{(*)+}}D^{^{-}}$  decays. As displayed in Fig. \ref{btoccdlp}
(a),  $\mathcal{B}(\bar{B}^0_{d}\to D^{^{+}}D^{^{-}})$ is  not very
sensitive with $|\lambda'^*_{i23}\lambda'_{i21}|$ and  has only
small allowed values when $|\phi_{\rm RPV}|$ is small. Fig.
\ref{btoccdlp} (b) shows that $\mathcal{B}(\bar{B}^0_{d}\to
D^{^{*+}}D^{^{-}})$ is decreasing
 with $|\lambda'^*_{i23}\lambda'_{i21}|$ and is weakly  sensitive to
 $|\phi_{\rm RPV}|$.
 Fig. \ref{btoccdlp} (c) exhibits the
slepton exchange coupling effects on
$\mathcal{C}(B^0_d,\bar{B}^0_d\to D^{+}D^{-})$, which is decreasing
with $|\lambda'^*_{i23}\lambda'_{i21}|$ and has little sensitivity
to $\phi_{\rm RPV}$. Slepton  exchange couplings have great effects
on $\mathcal{C}(B^0_d,\bar{B}^0_d\to
 D^{*+}D^{-},D^{+}D^{*-})$  and
$\mathcal{S}(B^0_d,\bar{B}^0_{d}\to
D^{^{+}}D^{^{-}},D^{^{*+}}D^{^{-}},D^{^{+}}D^{^{*-}})$, and they
have quite complex variational trends to
$|\lambda'^*_{i23}\lambda'_{i21}|$ and $|\phi_{\rm RPV}|$. The
effects of $\lambda'^*_{i23}\lambda'_{i21}$ couplings on
  $\mathcal{C}(B^0_d,\bar{B}^0_d\to
 D^{*+}D^{-})$ and
$\mathcal{S}(B^0_d,\bar{B}^0_{d}\to
D^{^{+}}D^{^{-}},D^{^{*+}}D^{^{-}})$ are shown in Fig.
\ref{btoccdlp} (d-f). $\mathcal{C}(B^0_d,\bar{B}^0_d\to
 D^{+}D^{*-})$ has  entirely different trends from $\mathcal{C}(B^0_d,\bar{B}^0_d\to
 D^{*+}D^{-})$. $\mathcal{S}(B^0_d,\bar{B}^0_d\to
 D^{+}D^{*-})$ has a similar trends as $\mathcal{S}(B^0_d,\bar{B}^0_d\to
 D^{*+}D^{-})$.

\subsection{Exclusive color-allowed  $b\to c\bar{c}s$ decays }

Exclusive color-allowed $b\to c\bar{c}s$ tree decays include
$\bar{B}^0_{d}\to D^{^{(*)+}}D^{^{(*)-}}_s$, $B^-_{u}\to
D^{^{(*)0}}D^{^{(*)-}}_s$  and $\bar{B}^0_{s}\to
D^{^{(*)+}}_sD^{^{(*)-}}_s$ decay modes. Almost all branching ratios
and one longitudinal polarization have been  measured by Belle
\cite{Zupanc:2007pu}, B{\footnotesize A}B{\footnotesize AR}
\cite{Aubert:2006nm,Aubert:2003jj,Aubert:2005xu}, CLEO
\cite{Gibaut:1995tu,Bortoletto:1991kz,Ahmed:2000ad,Bortoletto:1990fx},
and ARGUS \cite{Albrecht:1991pa} collaborations. Their averaged
values from Particle Data Group  \cite{PDG} are listed as follows
\begin{eqnarray}
\mathcal{B}(\bar{B}^0_d\to
D^{+}D^{-}_s)&=&(7.4\pm0.7)\times10^{-3},~~
\mathcal{B}(\bar{B}^0_d\to D^{*+}D^{-}_s)=(8.3\pm1.1)\times10^{-3},\nonumber\\
\mathcal{B}(\bar{B}^0_d\to
D^{+}D^{*-}_s)&=&(7.6\pm1.6)\times10^{-3},~~
\mathcal{B}(\bar{B}^0_d\to D^{*+}D^{*-}_s)=(17.9\pm1.4)\times10^{-3},\nonumber\\
\mathcal{B}(B^-_u\to D^{0}D^{-}_s)&=&(10.3\pm1.7)\times10^{-3},
~~\mathcal{B}(B^-_u\to D^{*0}D^{-}_s)=(8.4\pm1.7)\times10^{-3},\nonumber\\
\mathcal{B}(B^-_u\to D^{0}D^{*-}_s)&=&(7.8\pm1.6)\times10^{-3},~~
\mathcal{B}(B^-_u\to D^{*0}D^{*-}_s)=(17.5\pm2.3)\times10^{-3},\nonumber\\
\mathcal{B}(\bar{B}^0_s\to
D^{+}_sD^{-}_s)&=&(11\pm4)\times10^{-3},~~~~~
\mathcal{B}(\bar{B}^0_s\to D^{+}_sD^{*-}_s)<121\times10^{-3},\nonumber\\
\mathcal{B}(\bar{B}^0_s\to
D^{*+}_sD^{*-}_s)&<&257\times10^{-3},~~~~~~~~~~f_L (\bar{B}^0_d\to
D^{*+}D^{*-}_s)=0.52\pm0.05. \label{Eq:btoccsdata}
\end{eqnarray}

\begin{table}[tb]
\caption{\small Theoretical predictions for $CP$ averaged
$\mathcal{B}$ (in units of $10^{-4}$) and polarization fractions (in
units of $10^{-2}$) of exclusive color-allowed $b\to c\bar{c}s$
decays in the SM and the RPV MSSM.}
\begin{center}
\begin{tabular}{lccc}\hline\hline
~~~~Observable&SM &MSSM w/
$\lambda''^*_{231}\lambda''_{221}$~~~&MSSM w/
$\lambda'^*_{i23}\lambda'_{i22}$~~~
\\\hline $\mathcal{B}(\bar{B}^0_d\to D^+D^-_s)$&$[6.70,10.65]$&$[6.38,7.59]$&$[6.42,8.80]$\\\hline
$\mathcal{B}(\bar{B}^0_d\to
D^{*+}D^-_s)$&$[6.70,10.45]$&$[6.47,9.49]$&$[6.16,9.30]$\\\hline
$\mathcal{B}(\bar{B}^0_d\to
D^+D^{*-}_s)$&$[7.32,13.22]$&$[6.90,10.29]$&\\\hline
$\mathcal{B}(\bar{B}^0_d\to
D^{*+}D^{*-}_s)$&$[19.27,34.42]$&$[18.59,20.70]$&\\\hline
$\mathcal{B}(B^-_u\to
D^{0}D^{-}_s)$&$[7.21,11.43]$&$[6.90,8.12]$&$[6.90,9.49]$\\\hline
$\mathcal{B}(B^-_u\to
D^{*0}D^{-}_s)$&$[7.17,11.24]$&$[6.95,10.17]$&$[6.64,9.96]$\\\hline
$\mathcal{B}(B^-_u\to
D^{0}D^{*-}_s)$&$[7.89,14.27]$&$[7.43,11.00]$&\\\hline
$\mathcal{B}(B^-_u\to
D^{*0}D^{*-}_s)$&$[20.57,37.06]$&$[19.99,22.10]$&\\\hline
$\mathcal{B}(\bar{B}^0_s\to
D^+_sD^-_s)$&$[6.55,10.72]$&$[6.36,7.71]$&$[6.23,9.05]$\\\hline
$\mathcal{B}(\bar{B}^0_s\to
D^{*+}_sD^-_s)$&$[6.46,10.44]$&$[6.51,9.46]$&$[6.02,9.26]$\\\hline
$\mathcal{B}(\bar{B}^0_s\to
D^+_sD^{*-}_s)$&$[7.08,12.97]$&$[7.00,10.40]$&\\\hline
$\mathcal{B}(\bar{B}^0_s\to
D^{*+}_sD^{*-}_s)$&$[18.64,33.83]$&$[18.48,20.93]$&\\\hline
$f_L(\bar{B}^0_d\to
D^{*+}D^{*-}_s)$&$[50.25,50.91]$&$[48.46,51.13]$&\\\hline
$f_L(B^-_u\to
D^{*0}D^{*-}_s)$&$[50.28,50.94]$&$[48.49,51.16]$&\\\hline
$f_L(\bar{B}^0_s\to
D^{*+}_sD^{*-}_s)$&$[50.40,51.10]$&$[48.71,51.30]$&\\\hline
$f_\perp(\bar{B}^0_d\to
D^{*+}D^{*-}_s)$&$[8.85,9.55]$&$[8.15,12.85]$&\\\hline
$f_\perp(B^-_u\to
D^{*0}D^{*-}_s)$&$[8.87,9.57]$&$[8.17,12.88]$&\\\hline
$f_\perp(\bar{B}^0_s\to
D^{*+}_sD^{*-}_s)$&$[8.38,9.07]$&$[7.71,12.23]$&\\\hline \hline
\end{tabular}
\end{center}\label{Table:btoccsBRFL}
\end{table}

\begin{table}[tb]
\caption{\small  Theoretical predictions for CPAs (in units of
$10^{-2}$) of exclusive color-allowed $b\to c \bar c s$ decays in the SM
and the RPV MSSM.}
\begin{center}
\begin{tabular}{lccc}\hline\hline
~~~~Observable& SM &MSSM w/ $\lambda''^*_{231}\lambda''_{221}$&MSSM
w/ $\lambda'^*_{i23}\lambda'_{i22}$
\\\hline
$\mathcal{A}^{\rm dir}_{\rm CP}(\bar{B}^0_d\to
D^+D^-_s)$&$[-0.34,-0.22]$&$[-3.06,2.58]$&$[-8.42,7.94]$\\\hline
$\mathcal{A}^{\rm dir}_{\rm CP}(\bar{B}^0_d\to
D^{*+}D^-_s)$&$[0.03,0.06]$&$[-0.32,0.36]$&$[-0.98,1.07]$\\\hline
$\mathcal{A}^{\rm dir}_{\rm CP}(\bar{B}^0_d\to
D^+D^{*-}_s)$&$[-0.07,-0.06]$&$[-0.51,0.44]$&\\\hline
$\mathcal{A}^{+,{\rm dir}}_{\rm CP}(\bar{B}^0_d\to
D^{*+}D^{*-}_s)$&$[-0.07,-0.06]$&$[-0.69,0.56]$&\\\hline
$\mathcal{A}^{\rm dir}_{\rm CP}(B^-_u\to
D^0D^-_s)$&$[-0.34,-0.22]$&$[-3.06,2.58]$&$[-8.42,7.94]$\\\hline
$\mathcal{A}^{\rm dir}_{\rm CP}(B^-_u\to
D^{*0}D^-_s)$&$[0.03,0.06]$&$[-0.32,0.36]$&$[-0.98,1.07]$\\\hline
$\mathcal{A}^{\rm dir}_{\rm CP}(B^-_u\to
D^0D^{*-}_s)$&$[-0.07,-0.06]$&$[-0.51,0.58]$&\\\hline
$\mathcal{A}^{+,{\rm dir}}_{\rm CP}(B^-_u\to
D^{*0}D^{*-}_s)$&$[-0.07,-0.06]$&$[-0.69,0.56]$&\\\hline
$\mathcal{S}(B^0_s,\bar{B}^0_s\to
D^+_sD^-_s)$&$[0.40,0.61]$&$[-59.67,61.80]$&$[-99.84,99.79]$\\\hline
$\mathcal{S}(B^0_s,\bar{B}^0_s\to
D^{*+}_sD^-_s)$&$[1.33,2.16]$&$[-46.48,47.65]$&$[-56.04,59.14]$\\\hline
$\mathcal{S}(B^0_s,\bar{B}^0_s\to
D^+_sD^{*-}_s)$&$[-2.20,-1.31]$&$[-49.26,45.22]$&$[-58.96,56.60]$\\\hline
$\mathcal{S}^+(B^0_s,\bar{B}^0_s\to
D^{*+}_sD^{*-}_s)$&$[-31.81,-29.15]$&$[-54.41,55.49]$&\\\hline
$\mathcal{C}(B^0_s,\bar{B}^0_s\to
D^+_sD^-_s)$&$[0.22,0.34]$&$[-2.58,3.06]$&$[-7.94,8.42]$\\\hline
$\mathcal{C}(B^0_s,\bar{B}^0_s\to
D^{*+}_sD^-_s)$&$[2.77,13.39]$&$[3.15,10.13]$&$[0.79,28.22]$\\\hline
$\mathcal{C}(B^0_s,\bar{B}^0_s\to
D^+_sD^{*-}_s)$&$[-13.36,-2.76]$&$[-10.08,-3.14]$&$[-29.14,-0.54]$\\\hline
$\mathcal{C}^+(B^0_s,\bar{B}^0_s\to
D^{*+}_sD^{*-}_s)$&$[0.06,0.07]$&$[-0.56,0.69]$&\\\hline\hline
\end{tabular}
\end{center}\label{Table:btoccsACP}
\end{table}

 The SM predictions, in which the full theoretical uncertainties of input
parameters are considered, are given in the second  columns of Table
\ref{Table:btoccsBRFL} and Table \ref{Table:btoccsACP}. Theoretical
predictions for the branching ratios and the polarization fractions
are given in Table \ref{Table:btoccsBRFL}. Predicted CPAs are also
given in Table \ref{Table:btoccsACP}. Compared with the experimental
data, only the SM predictions of $\mathcal{B}(\bar{B}^0_d\to
D^{*+}D^{*-}_s,B^-_u\to D^{*0}D^{*-}_s)$ are slightly larger than
the corresponding experimental data given in Eq.
(\ref{Eq:btoccsdata}), and all the other branching ratios are
consistent with the data within $1\sigma$ error level. For the
color-allowed $b\to c\bar{c}s$ decays the penguin effects are doubly
Cabibbo-suppressed and, therefore, play a
significantly less pronounced role in CPAs. These CPAs have not been
measured yet. We obtain that
 all CPAs  are  expected to be very small (about
$10^{-3}$ or $10^{-4}$ order) in the SM except
$\mathcal{S}(B^0_s,\bar{B}^0_s\to D^{*+}_sD^{*-}_s)$ and
$\mathcal{C}(B^0_s,\bar{B}^0_s\to D^{*+}_sD^{-}_s,D^{+}_sD^{*-}_s)$.
There is no  obvious signature of $CP$ violation in $B_s\to
D^{*\pm}_sD^{\mp}_s$ decays since $\mathcal{C}(B^0_s,\bar{B}^0_s\to
D^{*+}_sD^{-}_s)\approx-\mathcal{C}(B^0_s,\bar{B}^0_s\to
D^{+}_sD^{*-}_s)$ and $\mathcal{S}(B^0_s,\bar{B}^0_s\to
D^{*+}_sD^{-}_s)\approx-\mathcal{S}(B^0_s,\bar{B}^0_s\to
D^{+}_sD^{*-}_s)$.

\begin{figure}[b]
\begin{center}
\begin{tabular}{c}
\includegraphics[scale=1]{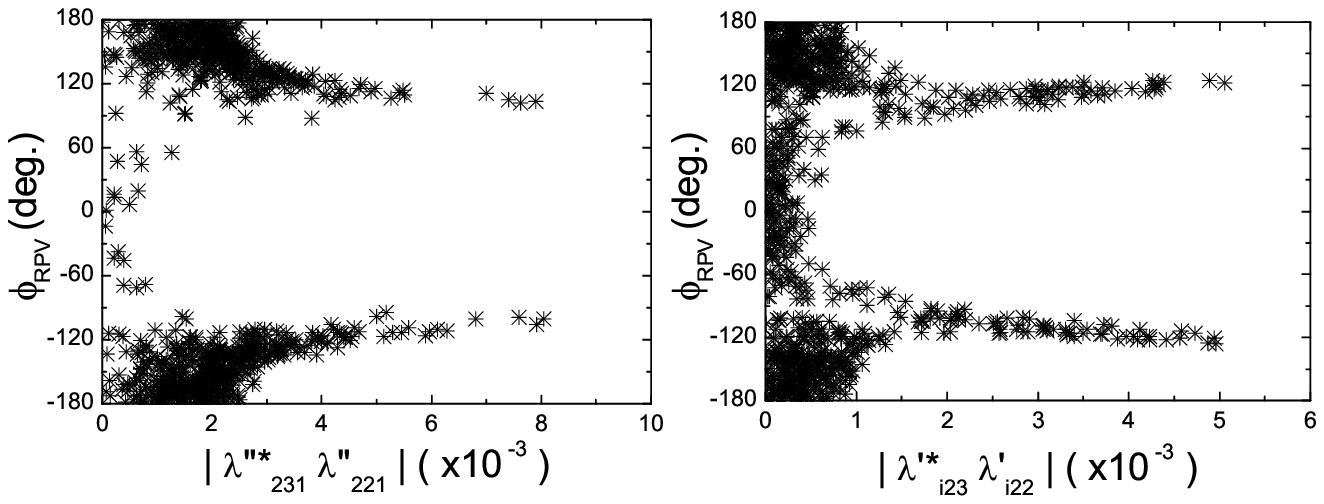}
\end{tabular}
\end{center}
\vspace{-0.6cm} \caption{\small Allowed parameter spaces for
relevant RPV coupling products constrained by the measurements of
exclusive color-allowed  $b\to c\bar{c}s$ decays listed in Eq.
(\ref{Eq:btoccsdata}).}
 \label{btoccsbounds}
\end{figure}

 There are two RPV coupling products,
$\lambda''^*_{231}\lambda''_{221}$ and
$\lambda'^*_{i23}\lambda'_{i22}$,  contributing to these exclusive
$b\to c \bar c s$ decay modes at tree level. We use the experimental data
listed  in Eq. (\ref{Eq:btoccsdata}) to constrain the RPV coupling
products, and the allowed spaces are shown in Fig.
\ref{btoccsbounds}.
The coupling $\lambda''^*_{231}\lambda''_{221}$ due to squark
exchange contributes to all twelve relative decay modes. The allowed
space of $\lambda''^*_{231}\lambda''_{221}$ is shown in the left
plot of Fig. \ref{btoccsbounds}.
 The slepton exchange couplings
$\lambda'^*_{i23}\lambda'_{i22}$ contribute to six $\bar{B}^0_{d}\to
D^{^{(*)+}}D^{^{-}}_s$, $B^-_{u}\to D^{^{(*)0}}D^{^{-}}_s$ and
$\bar{B}^0_{s}\to D^{^{(*)+}}_sD^{^{-}}_s$ decays, and the
constrained space is displayed in the right plot of Fig.
\ref{btoccsbounds}. {}From Fig. \ref{btoccsbounds}, we find both moduli of RPV couplings
have been limited as
$|\lambda''^*_{231}\lambda''_{221}| < 8.05\times 10^{-3}$  and
$|\lambda'^*_{i23}\lambda'_{i22}| < 5.05\times 10^{-3}$. Their
RPV weak phases are not constrained much when their magnitudes are
less than about $1\times 10^{-3}$.

Next, using the constrained  parameter spaces shown in Fig.
\ref{btoccsbounds},  we are going to predict RPV effects on the
observables which have not been measured yet.
 We summarize RPV MSSM
predictions with two separate RPV coupling contributions in the
last two columns of Table \ref{Table:btoccsBRFL} and Table
\ref{Table:btoccsACP}.

The contributions of $\lambda''^*_{231}\lambda''_{221}$ coupling due
to squark exchange are summarized  in the third columns of Table
\ref{Table:btoccsBRFL} and Table \ref{Table:btoccsACP}. In Table
\ref{Table:btoccsBRFL}, we find that the ranges of all branching
ratios are shrunk by $\lambda''^*_{231}\lambda''_{221}$ coupling and
the experimental constraints. Especially,
$\lambda''^*_{231}\lambda''_{221}$ coupling effects could
reduce the range of
 $\mathcal{B}(\bar{B}^0_{s}\to
D^{^{*+}}_sD^{^{*-}}_s)$.
However,
the allowed ranges of three
$f_L(B_{(s)}\to D^{*}_{(s)}D^{*}_s)$ and three $f_\perp(B_{(s)}\to
D^{*}_{(s)}D^{*}_s)$ are enlarged by
$\lambda''^*_{231}\lambda''_{221}$ coupling. In Table
\ref{Table:btoccsACP}, we can see $\lambda''^*_{231}\lambda''_{221}$
coupling does not affect
$\mathcal{C}(\bar{B}^0_{s}\to
D^{^{*+}}_sD^{^{-}}_s,D^{^{*+}}_sD^{^{-}}_s)$
much.

Meanwhile,  RPV coupling effects could remarkably
enlarge the allowed ranges of the other direct CPAs
(about 10 times).
Unfortunately, they are still too small to be measured
at presently available experiments.
It is interesting to note that  mixing-induced CPAs of $B_s$ decays are greatly
 affected by $\lambda''^*_{231}\lambda''_{221}$ coupling.
For an example,  $|\mathcal{S}(\bar{B}^0_{s}\to
D^{^{(*)+}}_sD^{^{(*)-}}_s)|$ could
be increased to   $\sim 50\%$  and their
signs could be changed by
the squark exchange coupling.

The contributions of $\lambda'^*_{i23}\lambda'_{i22}$ due to the
slepton exchanges are listed in the last columns of Table
\ref{Table:btoccsBRFL} and Table \ref{Table:btoccsACP}.  {}From Table
\ref{Table:btoccsBRFL}, we find the ranges of all branching ratios
are shrunk by $\lambda'^*_{i23}\lambda'_{i22}$ coupling and the
experimental constraints.  The last columns of Table
\ref{Table:btoccsACP} show that $\lambda'^*_{i23}\lambda'_{i22}$
coupling could enlarge
the ranges of all the CPAs.
Particularly,
$\lambda'^*_{i23}\lambda'_{i22}$ coupling could change the predicted
$\mathcal{S}(B^0_s,\bar{B}^0_{s}\to D^{^{(*)}}_sD_s)$ significantly
from quite narrow SM ranges to $[-0.6,0.6]$ or $[-1,1]$. The upper
limits of $|\mathcal{C}(B^0_s,\bar{B}^0_{s}\to
D^{^{*+}}_sD^{^{-}}_s,D^{^{+}}_sD^{^{*-}}_s)|$ are increased a lot
by $\lambda'^*_{i23}\lambda'_{i22}$ couplings.

\begin{figure}[t]
\begin{center}
\begin{tabular}{c}
\includegraphics[scale=1]{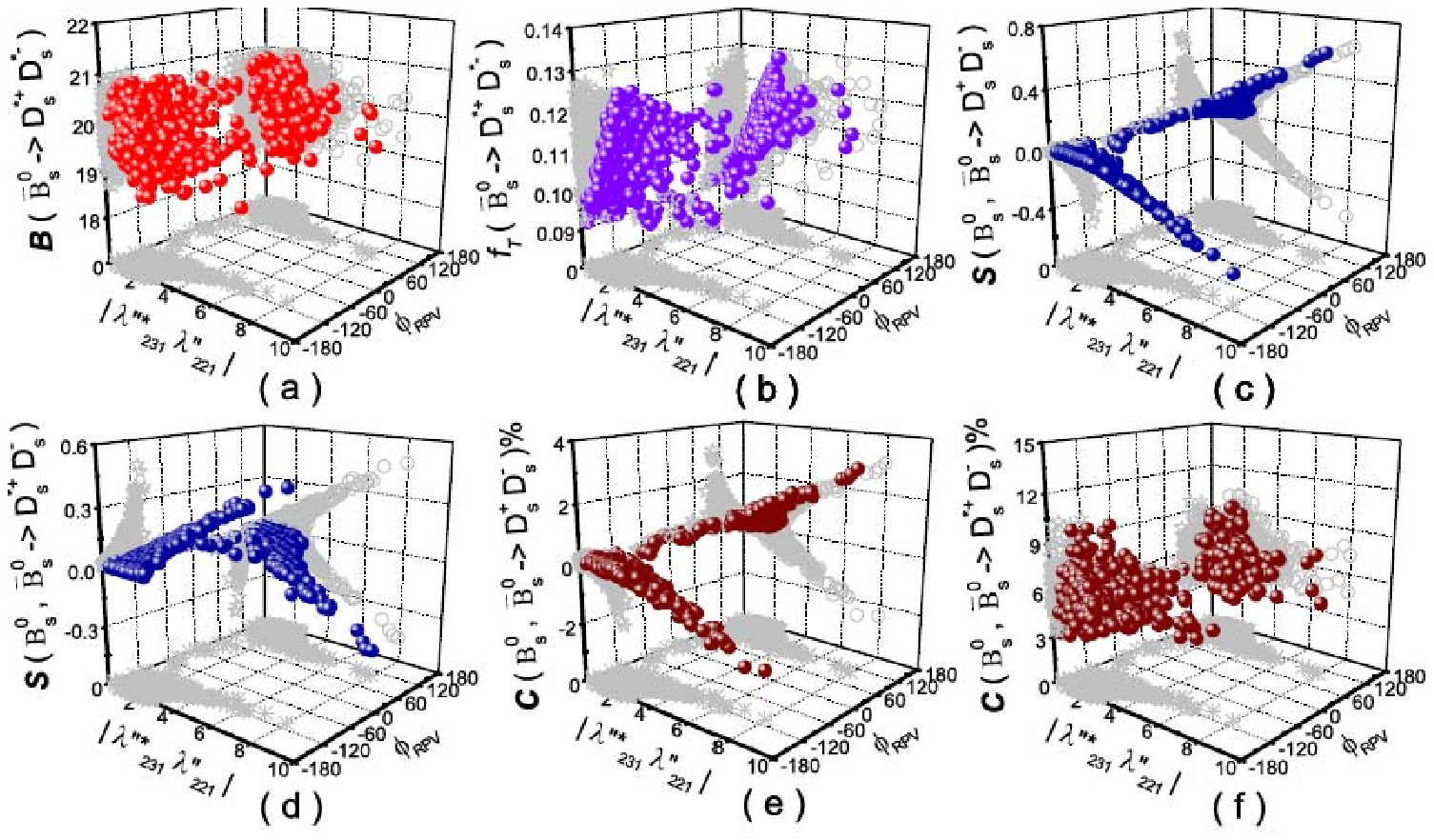}
\end{tabular}
\end{center}
\vspace{-0.6cm} \caption{\small Effects of RPV coupling
$\lambda''^*_{231}\lambda''_{221}$  in $\bar{B}^0_{s}\to
D_{s}^{^{(*)+}}D^{^{(*)-}}_s$
decays, where $|\lambda''^*_{231}\lambda''_{221}|$ is in units of
$10^{-3}$, $\mathcal{B}$ in units of $10^{-4}$, and $f_T$ denotes the transverse polarization fraction
$f_\perp$.} \label{btoccslpp}
\vspace{1cm}
\begin{center}
\begin{tabular}{c}
\includegraphics[scale=1]{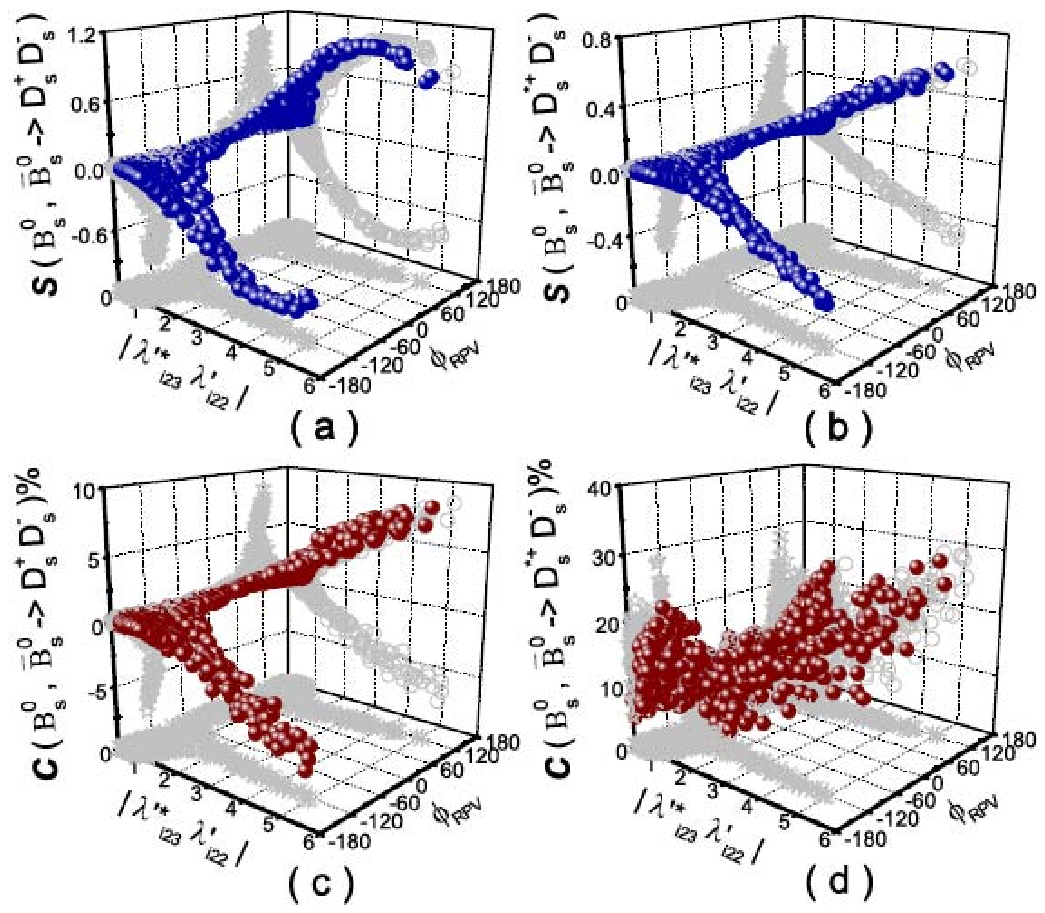}
\end{tabular}
\end{center}
\vspace{-0.6cm} \caption{\small Effects of
$\lambda'^*_{i23}\lambda'_{i22}$  in $\bar{B}^0_{s}\to
D_{s}^{^{(*)+}}D^{^{-}}_s$ decays, where
$|\lambda'^*_{i23}\lambda'_{i22}|$ are in units of $10^{-3}$, and $\mathcal{B}$ in units of $10^{-4}$.}
\label{btoccslp}
\end{figure}

Since RPV contributions to physical observables are also very
similar in $\bar{B}^0_{d}\to D^{^{(*)+}}D^{^{(*)-}}_s$, $B^-_{u}\to
D^{^{(*)0}}D^{^{(*)-}}_s$ and $\bar{B}^0_{s}\to
D^{^{(*)+}}_sD^{^{(*)-}}_s$ systems, we  show only a few  observables
of $B_s$ decays as examples. Fig. \ref{btoccslpp} and Fig.
\ref{btoccslp} show the variational trends in some observables
with the $\lambda''^*_{231}\lambda''_{221}$ and
$\lambda'^*_{i23}\lambda'_{i22}$ couplings, respectively.

First, we will elucidate the information implied in  Fig.
\ref{btoccslpp}.
{}From Fig. \ref{btoccslpp} (a), we find $\mathcal{B}(\bar{B}^0_s\to D^{*+}_sD^{*-}_s)$
is not changed much by
$|\lambda''^*_{231}\lambda''_{221}|$, and could have only small
value when $\phi_{\rm RPV}$ is not too large.
As shown in Fig. \ref{btoccslpp}(b),
$f_{\perp}(\bar{B}^0_s\to D^{*+}D^{*-}_s)$ is increasing with
$|\lambda''^*_{231}\lambda''_{221}|$ and also could have small
value when $\phi_{\rm RPV}$ is small.
 {}From Fig. \ref{btoccslpp} (c-d), we find
$|\mathcal{S}(B^0_s,\bar{B}^0_{s}\to
D^{^{+}}_sD^{^{-}}_s,D^{^{*+}}_sD^{^{-}}_s)|$ are  all
rapidly  increasing with $|\lambda''^*_{231}\lambda''_{221}|$ and could be
very large at the large values of
$|\lambda''^*_{231}\lambda''_{221}|$ and $|\phi_{\rm RPV}|$,
furthermore, the RPV weak phase has opposite effects between
$|\mathcal{S}(B^0_s,\bar{B}^0_{s}\to D^{^{+}}_sD^{^{-}}_s)|$ and
$|\mathcal{S}(B^0_s,\bar{B}^0_{s}\to D^{^{*+}}_sD^{^{-}}_s)|$.
$\lambda''^*_{231}\lambda''_{221}$ coupling effects on
$\mathcal{S}(B^0_s,\bar{B}^0_s\to
 D^{+}_sD^{*-}_s)$ ($\mathcal{S}(B^0_s,\bar{B}^0_s\to
 D^{*+}_sD^{*-}_s)$) are similar as ones on  $\mathcal{S}(B^0_s,\bar{B}^0_s\to
 D^{*+}_sD^{-}_s)$ ($\mathcal{S}(B^0_s,\bar{B}^0_s\to
 D^{+}_sD^{-}_s)$).
 So any measurement of $\mathcal{S}(B^0_s,\bar{B}^0_{s}\to
D^{^{(*)+}}_sD^{^{(*)-}}_s)$ in the future will strongly constrain
the magnitude and RPV weak phase of
$\lambda''^*_{231}\lambda''_{221}$ coupling, and then other
mixing-induced CPAs will be more accurately predicted as indicated
by Fig. \ref{btoccslpp} (c-d). As shown in Fig. \ref{btoccslpp} (e),
$\mathcal{C}(B^0_s,\bar{B}^0_{s}\to D^{^{+}}_sD^{^{-}}_s)$ has
similar trends as $\mathcal{S}(B^0_s,\bar{B}^0_{s}\to
D^{^{+}}_sD^{^{-}}_s)$ with $|\lambda''^*_{231}\lambda''_{221}|$ and
$\phi_{\rm RPV}$,  however, $\lambda''^*_{231}\lambda''_{221}$
coupling effects on the former are much smaller than the effects on
the latter. $\mathcal{A}^{\rm dir}_{\rm CP}(B_{u,d}\to
DD^{^{-}}_s,D^{^{*}}D^{^{-}}_s,D^{^{*}}D^{^{*-}}_s)$ and
$-\mathcal{A}^{\rm dir}_{\rm CP}(B_{u,d}\to DD^{^{*-}}_s)$ have the
same variational trends with $\lambda''^*_{231}\lambda''_{221}$ as
$\mathcal{C}(B^0_s,\bar{B}^0_{s}\to D^{^{+}}_sD^{^{-}}_s)$ has.
$\mathcal{C}(B^0_s,\bar{B}^0_{s}\to
D^{^{*+}}_sD^{^{-}}_s,D^{^{*+}}_sD^{^{-}}_s)$ are not affected much
by  $\lambda''^*_{231}\lambda''_{221}$ coupling, and we show
$\mathcal{C}(B^0_s,\bar{B}^0_{s}\to D^{^{*+}}_sD^{^{-}}_s)$ in Fig.
\ref{btoccslpp} (f) as an example.

Fig. \ref{btoccslp} illustrates $\lambda'^*_{i23}\lambda'_{i22}$
contributions to the CPAs of $\bar{B}^0_{s}\to
D^{^{(*)+}}_sD^{^{-}}_s$. As displayed in Fig. \ref{btoccslp} (a-b),
$\mathcal{S}(B^0_s,\bar{B}^0_{s}\to D^{^{(*)+}}_sD^{^{-}}_s)$ are
very sensitive to $\lambda'^*_{i23}\lambda'_{i22}$ couplings.
$|\mathcal{S}(B^0_s,\bar{B}^0_{s}\to D^{^{(*)+}}_sD^{^{-}}_s)|$ are
strongly increasing with $|\lambda'^*_{i23}\lambda'_{i22}|$, and
they could reach extremum at $|\phi_{\rm RPV}|\approx 120^\circ$.
The $\lambda'^*_{i23}\lambda'_{i22}$ coupling effects on
$\mathcal{S}(B^0_s,\bar{B}^0_{s}\to D^{^{+}}_sD^{^{*-}}_s)$ are same
as the ones on $\mathcal{S}(B^0_s,\bar{B}^0_{s}\to
D^{^{*+}}_sD^{^{-}}_s)$.
 Fig. \ref{btoccslp} (c) shows that
$\mathcal{C}(B^0_s,\bar{B}^0_{s}\to D^{^{+}}_sD^{^{-}}_s)$  are also
very sensitive to
$|\lambda'^*_{i23}\lambda'_{i22}|$
and $\phi_{\rm RPV}$, but it is
still too small to be measured
in near future.
In addition, $\mathcal{A}^{\rm dir}_{\rm CP}(B_{u,d}\to
D^{^{(*)}}D^{^{-}}_s)$ are
affected much
by $\lambda'^*_{i23}\lambda'_{i22}$ couplings, and just RPV MSSM
predictions of these quantities are very small.
$\lambda'^*_{i23}\lambda'_{i22}$ couplings could have similar
impacts on $\mathcal{C}(B^0_s,\bar{B}^0_{s}\to
D^{^{*+}}_sD^{^{-}}_s)$ and $-\mathcal{C}(B^0_s,\bar{B}^0_{s}\to
D^{^{*+}}_sD^{^{-}}_s)$. We give these coupling effects on
$\mathcal{C}(B^0_s,\bar{B}^0_{s}\to D^{^{*+}}_sD^{^{-}}_s)$ in Fig.
\ref{btoccslp} (d),
which shows
 $\mathcal{C}(B^0_s,\bar{B}^0_{s}\to
D^{^{*+}}_sD^{^{-}}_s)$ is increasing with
$|\lambda'^*_{i23}\lambda'_{i22}|$, and could have large value at
large $|\phi_{\rm RPV}|$.

\section{Summary}

We have studied the twenty-four  double charm decays
$\bar{B}^0_{d}\to D^{^{(*)+}}D^{^{(*)-}}_{_{(s)}}$,  $B^-_{u}\to
D^{^{(*)0}}D^{^{(*)-}}_{_{(s)}}$ and $\bar{B}^0_{s}\to
D^{^{(*)+}}_sD^{^{(*)-}}_{_{(s)}}$  in the RPV MSSM. We have treated
these decays in the naive factorization and
removed the known $k^{2}$ ambiguities in the penguin contributions via $b\to q g^{*}(\gamma^{*}) \to q c{\bar c}$
by calculating its  hard kernel $b\to c +D_{q}$.
Considering the theoretical
uncertainties and the experimental error-bars, we have obtained fairly
constrained parameter spaces of RPV couplings from the present
experimental data, which have quite highly consistent measurements
among the relative collaborations. Furthermore, using the
constrained RPV coupling parameter spaces, we have predicted the RPV
effects on the branching ratios, the CPAs and the polarization
fractions, which have not been measured or have not been well
measured yet.

The investigation of exclusive color-allowed $b\to c\bar{c}d$ decays
is motivated by the large direct CPA $\mathcal{C}(B^0_d, \bar{B}^0_d\to D^+D^-)$ reported by Belle,
which has not been confirmed by B{\footnotesize A}B{\footnotesize AR} yet and contradicted
the SM prediction.
 Using the most conservative experimental bounds
from $\bar{B}^0_{d}\to D^{^{(*)+}}D^{^{(*)-}}$ and $B^-_{u}\to
D^{^{(*)0}}D^{^{(*)-}}$ systems (choose only twelve highly consistent
measurements between B{\footnotesize A}B{\footnotesize AR} and
Belle), we have first obtained quite strong constraints on the
involved RPV couplings $\lambda''^*_{232}\lambda''_{212}$ and
$\lambda'^*_{i23}\lambda'_{i21}$ from $b\to c\bar{c}d$ transition,
due to squark exchange and slepton exchanges, respectively.
Then, using the constrained RPV coupling parameter spaces, we have
predicted the RPV effects on $\mathcal{C}(B^0_d, \bar{B}^0_d\to
D^+D^-)$ and other observables, which have less consistent
measurements  or have not been  measured yet.
We have found that the lower limit of
$\mathcal{C}(B^0_d,\bar{B}^0_d\to D^{(*)+}D^{-})$ could be slightly
reduced by the RPV couplings. Our RPV MSSM prediction of
$\mathcal{C}(B^0_d,\bar{B}^0_{d}\to D^{^{+}}D^{^{-}})$ is consistent
with B{\footnotesize A}B{\footnotesize AR} measurement within $1
\sigma$ error level, but cannot explain
the corresponding Belle
experimental data within $3 \sigma$
level. We have also found
that  the contributions of $\lambda''^*_{232}\lambda''_{212}$ and
$\lambda'^*_{i23}\lambda'_{i21}$ cannot affect the relevant
branching ratios much.
$\lambda''^*_{232}\lambda''_{212}$ or
$\lambda'^*_{i23}\lambda'_{i21}$ contributions could greatly enlarge
the ranges of the relevant mixing-induced CPAs
$\mathcal{S}(B^0_d,\bar{B}^0_{d}\to D^{^{(*)+}}D^{^{(*)-}})$ from
their SM predictions, and these quantities are very sensitive to the
moduli and weak phases of $\lambda''^*_{232}\lambda''_{212}$ and
$\lambda'^*_{i23}\lambda'_{i21}$. So more accurate measurements of
$\mathcal{S}(B^0_d,\bar{B}^0_{d}\to D^{^{(*)+}}D^{^{(*)-}})$ in
the future will much more strongly constrain these RPV couplings, and then
mixing-induced CPAs can be more accurately predicted as well.
Effects of $\lambda''^*_{232}\lambda''_{212}$ coupling could
slightly extend the allowed regions of four
$\mathcal{C}(B^0_d,\bar{B}^0_d\to D^{(*)+}D^{(*)-})$ and eight
$\mathcal{A}^{\rm dir}_{\rm CP}(B^-_{u}\to
D^{(*)0}D^{(*)-},\bar{B}^0_{s}\to D^{(*)+}_sD^{(*)-})$.
$\mathcal{C}(B^0_d,\bar{B}^0_d\to D^{(*)+}D^{-})$ are also sensitive
to the slepton exchange couplings $\lambda'^*_{i23}\lambda'_{i21}$,
and their signs could be changed by these couplings.
Additionally,
 three $f_L(B_{(s)}\to D^{*}_{(s)}D^{*})$ and three
$f_\perp(B_{(s)}\to D^{*}_{(s)}D^{*})$ are decreased and increased
by $\lambda''^*_{232}\lambda''_{212}$ coupling, respectively, and
their allowed ranges are magnified by these couplings.

For  $\bar{B}^0_{d}\to D^{^{(*)+}}D^{^{(*)-}}_s$, $B^-_{u}\to
D^{^{(*)0}}D^{^{(*)-}}_s$ and $\bar{B}^0_{s}\to
D^{^{(*)+}}_sD^{^{(*)-}}_s$ decays, RPV couplings
$\lambda''^*_{231}\lambda''_{221}$ and
$\lambda'^*_{i23}\lambda'_{i22}$  contribute to these decay modes.
We have found $\lambda''^*_{231}\lambda''_{221}$ coupling
effects could apparently suppress the upper limit of
$\mathcal{B}(\bar{B}^0_{s}\to D^{^{*+}}_sD^{^{*-}}_s)$, and could
slightly enlarge the allowed ranges of three $f_L(B_{(s)}\to
D^{*}_{(s)}D^{*}_s)$ and three $f_\perp(B_{(s)}\to
D^{*}_{(s)}D^{*}_s)$, nevertheless these quantities are not very
sensitive to the changes of $|\lambda''^*_{231}\lambda''_{221}|$ and
$\phi_{\rm RPV}$.
 $\mathcal{C}(\bar{B}^0_{s}\to
D^{^{*+}}_sD^{^{-}}_s,D^{^{+}}_sD^{^{*-}}_s)$ are not evidently
affected by the squark exchange $\lambda''^*_{231}\lambda''_{221}$
coupling, and their upper limits are increased a lot by the slepton
exchange $\lambda'^*_{i23}\lambda'_{i22}$ couplings. RPV couplings
$\lambda''^*_{231}\lambda''_{221}$ and
$\lambda'^*_{i23}\lambda'_{i22}$ could greatly enlarge all other
$CP$ asymmetries, which are also very sensitive to the relevant RPV
couplings. However, the direct CPAs are still too small to be
measured soon. We could explore RPV MSSM effects from the
mixing-induced CPAs of $B_s$ decays.

With the large amount of B decay data from BABAR and Belle,
especially from LHCb in the near future, measurements of previously
known observables will become more precise and many unobserved
observables will be also measured. {}From the comparison of our
predictions in Figs. \ref{btoccdlpp}-\ref{btoccdlp} and Figs
\ref{btoccslpp}-\ref{btoccslp} with near future experiments, one
will obtain more stringent bounds on the product combinations of the
RPV couplings. On the other hand, the RPV MSSM predictions of other
decays will become more precise by the more stringent bounds on the
RPV couplings.
 The results in this paper could be useful for probing RPV MSSM
 effects, and will correlate  with searches for direct supersymmetry  signals at
future experiments, for example, the LHC.

\section*{Acknowledgments}

The work of C.S. Kim was supported in part by CHEP-SRC and in part
by the KRF Grant funded by the Korean Government (MOEHRD) No.
KRF-2005-070-C00030.  The work of Ru-Min Wang was supported by the
second stage of Brain Korea 21 Project. The work of Ya-Dong Yang was supported by the National
Science Foundation under contract Nos. 10675039 and 10735080.

\end{document}